\documentclass[final, 5p, times]{elsarticle}
\usepackage[T1]{fontenc}
\usepackage[latin9]{inputenc}
\usepackage{amsmath}
\usepackage{amssymb}
\usepackage{graphicx}
\usepackage{esint}
\usepackage[caption=false, font=footnotesize]{subfig}

\usepackage{booktabs}

\journal{NIMA}
\usepackage{siunitx}[=v2]
\usepackage{upgreek}
\usepackage[bookmarks,bookmarksopen,bookmarksdepth=6]{hyperref}
\hypersetup{
    colorlinks=true,
    linkcolor=blue,
    filecolor=magenta,      
    urlcolor=cyan
}

\usepackage{cases}
\usepackage{url}

\biboptions{sort&compress}
\begin{document}

\begin{frontmatter}

\title{Investigation of high resistivity $p$-type
FZ silicon diodes after $^\text{60}$Co $\gamma$-irradiation}
\tnotetext[mytitlenote]{Work performed in the frame of the
\href{https://rd50.web.cern.ch/}{CERN-RD50} collaboration.}

\author[mymainaddress]{C.~Liao\corref{mycorrespondingauthor}}
\cortext[mycorrespondingauthor]{Corresponding author}
\ead{chuan.liao@desy.de}

\author[mymainaddress]{E.~Fretwurst}
\author[mymainaddress]{E.~Garutti}
\author[mymainaddress]{J.~Schwandt}
\author[mysecondaryaddress]{I.~Pintilie}
\author[mythirdaddress]{A.~Himmerlich}
\author[mythirdaddress]{M.~Moll}
\author[mythirdaddress]{Y.~Gurimskaya}
\author[myfourthaddress]{Z.~Li}

\address[mymainaddress]{Institute for Experimental Physics, University of
Hamburg, Hamburg, Germany}
\address[mysecondaryaddress]{National Institute of Materials Physics, Bucharest,
Romania}
\address[mythirdaddress]{European Organization for Nuclear Research (CERN),
Geneva, Switzerland}
\address[myfourthaddress]{College of physics and optoelectronic engineering,
Ludong University, Yantai, China}

\begin{abstract}
In this work, the effects of $^\text{60}$Co $\gamma$-ray irradiation on high resistivity $p$-type diodes have been investigated. The diodes were exposed to dose values of 0.1, 0.2, 1, and \SI{2}{\mega Gy}. Both macroscopic ($I$--$V$, $C$--$V$) and microscopic (Thermally Stimulated Current~(TSC)) measurements were conducted to characterize the radiation-induced changes. The investigated diodes were manufactured on high resistivity $p$-type Float Zone (FZ) silicon and were further classified into two types based on the isolation technique between the pad and guard ring: $p$-stop and $p$-spray. After irradiation, the macroscopic results of current-voltage and capacitance-voltage measurements were obtained and compared with existing literature data. Additionally, the microscopic measurements focused on the development of the concentration of different radiation-induced defects, including the boron interstitial and oxygen interstitial (B$_\text{i}$O$_\text{i}$) complex, the carbon interstitial and oxygen interstitial C$_\text{i}$O$_\text{i}$ defect, the H40K, and the so-called I$_\text{P}^*$.
\par
To investigate the thermal stability of induced defects in the bulk, isochronal annealing studies were performed in the temperature range of \SI{80}{\celsius} to \SI{300}{\celsius}. These annealing processes were carried out on diodes irradiated with doses of 1 and \SI{2}{\mega Gy} and the corresponding TSC spectra were analysed. Furthermore, in order to investigate the unexpected results observed in the $C$-$V$ measurements after irradiation with high dose values, the surface conductance between the pad and guard ring was measured as a function of both dose and annealing temperature.
\end{abstract}


\end{frontmatter}

\section{Introduction}
In High Luminosity Large Hadron Collider (HL-LHC) experiments, strip and pixel silicon sensors in the inner tracking detectors have to cope with extraordinarily high particle rates of up to 200 p--p collisions per bunch crossing. New types of sensors to be used for this purpose, started to be developed. They are manufactured on boron doped ($p$-type) silicon and have different structures, from the simplest p-i-n diode to devices which amplifies the signals and have better time resolution, as e.g.\ the Low Gain Avalanche Detectors (LGADs)~\cite{b1,b2}. The degradation in the performance of these sensors is caused by the generation of electrically active radiation-induced defects. For instance, the charged defects at room temperature (e.g.\ E30K, B$_\text{i}$O$_\text{i}$, H140K, H152K and I$_\text{P}$)~\cite{b3, b4, b5, b22} cause changes of the space charge density ($N_\text{eff}$). Vacancy related traps such as V$_2$ and V$_3$~\cite{b6} lead to the degradation of charge collection efficiency (CCE) and an increase of dark current. In general, the radiation-induced defects can be classified according to their size: point defects, up to 2$\sim$3 atoms size, or more extended ones, also known as cluster-related defects. All the radiation-induced defects contribute to the degradation of silicon sensors and can be characterized using spectroscopic techniques such as Deep Level Transient Spectroscopy (DLTS), for low defect concentration, and Thermally Stimulated Current (TSC) after high levels of irradiation when DLTS does not work anymore. During a temperature scan, both, point and cluster-related defects are contributing to the signals observed in TSC and DLTS measurements. Detecting and analyzing the combined effects of these defects pose a challenge for these techniques. 
\par
The main radiation damage effect seen in $p$-type silicon sensors is the deactivation of the boron dopant. Thus, the atoms of substitutional Boron (B$_\text{s}$) may switch the sites with interstitial silicon (Si$_\text{i}$) created during radiation via a Watkins replacement mechanism~\cite{b33} and transform in Boron interstitial (B$_\text{i}$), losing its acceptor character. The B$_\text{i}$ atoms migrate in the crystal and react with interstitial Oxygen, found in abundance in silicon, leading to the formation of B$_\text{i}$O$_\text{i}$ complex. Another possible acceptor removal mechanism in $p$-type silicon considers the interaction between the negatively charged acceptor dopant (B$_\text{s}$ for boron-doped silicon) and the positively charged Si$_\text{i}$ atoms forming the B$_\text{s}$-Si$_\text{i}$ complex~\cite{b38, b39}. The boron removal effect induced by 23~GeV protons, 5.5~MeV electrons and 1~MeV neutrons was investigated in the framework of the CERN-RD50 "Acceptor removal project"~\cite{b7, b8, b9, b10, b11}. This effect can also be introduced by high-energy photons i.e.\ $^{60}$Co $\gamma$--rays. While hadrons can produce both point and cluster-related defects, $\gamma$--rays introduce only point defects~\cite{b12, b13}. In this work, $n^\text{+}$-$p$ diodes irradiated with $^{60}$Co $\gamma$--rays are investigated. This means that the changes in the macroscopic properties of silicon sensors, as determined from Current-Voltage ($I$--$V$)/Capacitance-Voltage ($C$--$V$) characteristics are attributed to point defects only. 
\par
The paper is structured as follows. Section~\ref{sec:2} includes information on the fabrication of the investigated diodes, irradiation details and measurement procedures. In section~\ref{sec:3} the results of the macroscopic ($I$--$V$/$C$--$V$) and microscopic (TSC) investigations, on as-irradiated and annealed samples, are presented and discussed. The conclusions of this work are presented in section~\ref{sec:4}.

\section{Experimental Details}\label{sec:2}

\begin{figure}[htbp]
  \centering
    \subfloat[\label{fig10a}]{
      \includegraphics[width=0.5\linewidth]{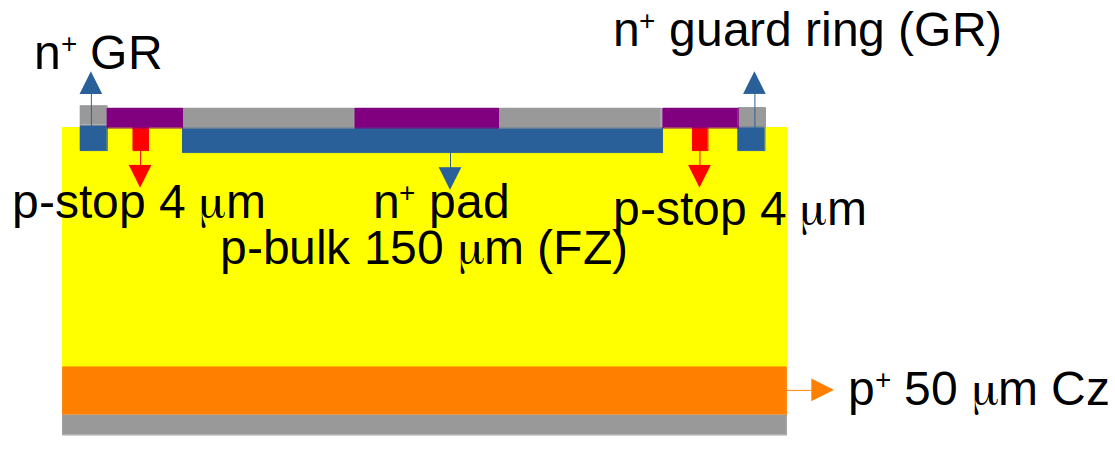}}
     \subfloat[\label{fig10b}]{
      \includegraphics[width=0.5\linewidth]{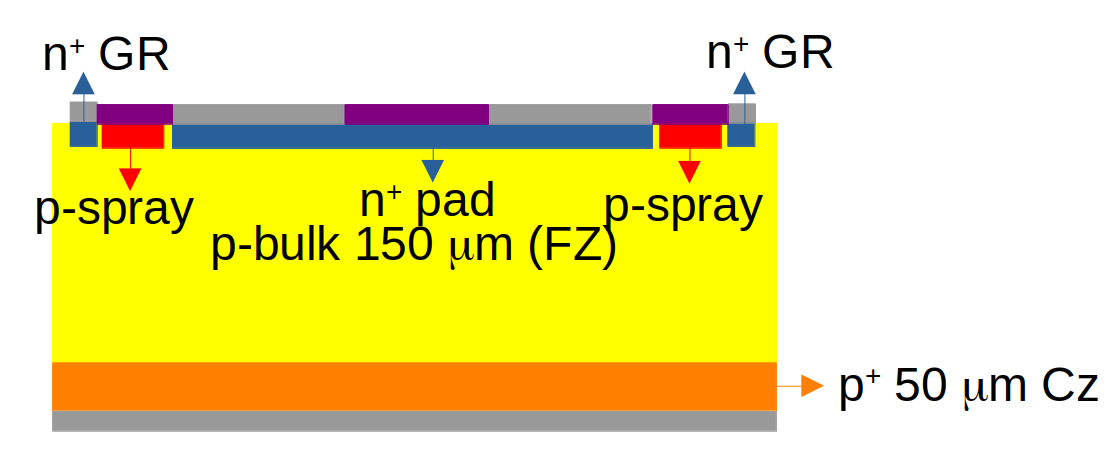}}\\
     \subfloat[\label{fig10c}]{
      \includegraphics[width=0.5\linewidth]{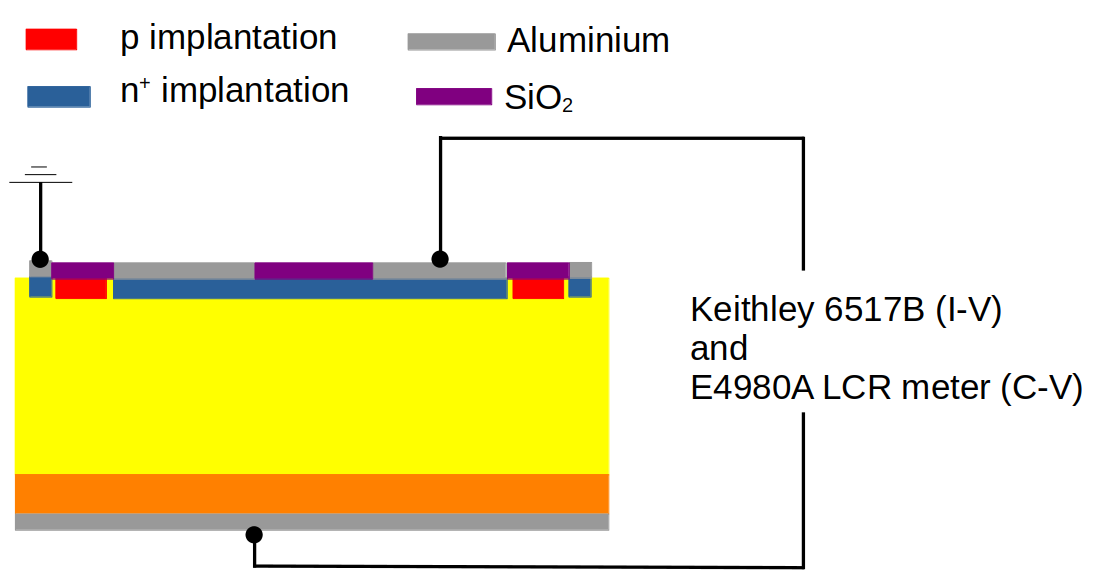}}
     \subfloat[\label{fig10d}]{
      \includegraphics[width=0.5\linewidth]{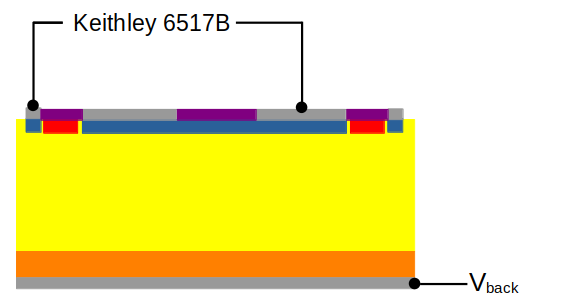}
    }
  \caption{(a) Side view of p-stop diodes. (b) Side view of p-spray diodes. (c) Schematic representation of the electrical setup used for $I$--$V$/$C$--$V$ measurements. (d) Schematic representation of the electrical circuit used for measuring the surface current.}
  \label{fig10}
\end{figure}

The study was performed on 12 $p$-type silicon diodes which were designed for upgrading the planar pixel sensors in the Compact Muon Solenoid (CMS) inner tracker for HL-LHC at CERN. These diodes were manufactured by Hamamatsu Photonics K. K. (HPK)~\cite{b40}, and were processed on, so-called, FDB150 (float zone Si-Si direct bonded) wafers, obtained by bonding together two wafers: a high resistivity float zone (FZ) wafer and a low resistivity handle one, produced with the Czochralski (Cz) method. The FZ wafer is thinned down to an active thickness of \SI{150}{\micro m} and after processing the pixel sensors, diodes and other test structures, the handle wafer is thinned down to \SI{50}{\micro m}, resulting in a total thickness of \SI{200}{\micro m}-see Fig.~\ref{fig10a} and Fig.~\ref{fig10b}.
\par
The initial boron doping of the FZ bulk is $N_\text{eff,0}$ = $\num{3.5e12}$~$\si{cm^{-3}}$, corresponding to a resistivity of about 4~k$\si{\Omega cm}$. Besides boron, the main impurity in the bulk is oxygen, having a concentration of about $\num{1e17}$~$\si{cm^{-3}}$~\cite{b14}. The backside Cz material has a boron content of $\num{1e19}$~$\si{cm^{-3}}$. According to the design of the isolation between the pad and guard ring structure, there are two types of diodes: with $p$-stop (FDB150P, simplified to F150P in this work)and $p$-spray (FDB150Y, simplified to F150Y in this work). The pad and guard ring are isolated by a thin (\SI{4}{\micro \meter}) boron-implanted region on the surface in the case of $p$-stop, while $p$-spray is a boron implant across the full wafer surface before the processing of the sensors-see Fig.~\ref{fig10a} and Fig.~\ref{fig10b}. The active area of the diodes is 0.25~$\si{cm^2}$. More details can be found in~\cite{b14, b15}.
\par
The 12 diodes were divided into four groups, each containing two p-stop type and one $p$-spray type diode. The proposed dose values for $^{60}$Co $\gamma$ irradiation of each of the groups were 0.1, 0.2, 1 and 2~MGy. The irradiation was performed using the $^{60}$Co Panoramic Irradiation facility at Rudjer Boskovic Institute \cite{b16}. Considering the radiation flux and exposure time, the following dose values were achieved: 94$\pm$0.96, 189$\pm$3.9, 924$\pm$27 and 1861$\pm$56~kGy. The investigated diodes and the achieved irradiation doses are given in Table~\ref{tab:table1}.
\begin{table*}[htbp]
\centering
\caption{Device information}
  \begin{tabular}{@{}lccccc@{}}
   \toprule
   $p$-stop1 & F150P-1 & F150P-3 & F150P-5 & F150P-8 \\
   $p$-stop2 & F150P-2 & F150P-4 & F150P-7 & F150P-9 \\
   $p$-spray & F150Y-1 & F150Y-2 & F150Y-5 & F150Y-8 \\
  \midrule
  Dose value ($\si{kGy}$) & 94$\pm$0.96 & 189$\pm$3.9 & 924$\pm$27 & 
  1861$\pm$56\\
  \bottomrule
  \end{tabular}
\label{tab:table1}
\end{table*} 

\par
The electrical performance of the diodes was measured by means of $I$--$V$ and $C$--$V$ characteristics at 20~$\pm$~\SI{0.01}{\celsius}. For all the diodes, the space charge density ($N_\text{eff}$) and the depleted depth $w(V)$ were determined from $C$--$V$ measurements corresponding to a frequency of 500~kHz. and an AC voltage of 500~mV. The electrical circuit for measuring I-V/C-V characteristics is depicted in Fig.~\ref{fig10c}. The current flow between the pad and the guarding was measured using the circuit given in Fig.~\ref{fig10d}. The radiation-induced electrically active defects were investigated by means of the TSC technique. Details of the experimental setup can be found in~\cite{b8,b13,b17,b18}. In this work, the trap filling was performed at different filling temperatures ($T_\text{fill}$ = 10$\sim$100~K, in steps of 10~K) by injection of a forward current ($I_\text{fill}$) produced by a forward bias of $V_\text{fill}$~=~10~V for 30 s. For $T_\text{fill}$ = 10~K, the $I_\text{fill}$ $\lesssim$ \SI{1}{\milli\ampere} and shows negligible dependence on the dose and the type of isolation. For $T_\text{fill}$ $\geq$ 20~K, all the $I_\text{fill}$ roughly equal to \SI{1}{\milli\ampere}. The TSC spectra were then recorded for different reverse biases $V_\text{bias}$ applied to the sample (-100 to -300~V in steps of 50~V) during heating up the samples with a constant heating rate of $\beta$ = 0.183~K/s. Isochronal annealing experiments were conducted using a Thermo SCIENTIFIC oven~\cite{b19} with air as the annealing atmosphere.

\section{Results and Discussion}\label{sec:3}
\subsection{As-irradiated devices} \label{sec:3.1}

\begin{figure}[htbp]
  \centering
    \subfloat[\label{fig1a}]{
      \includegraphics[width=1.0\linewidth]{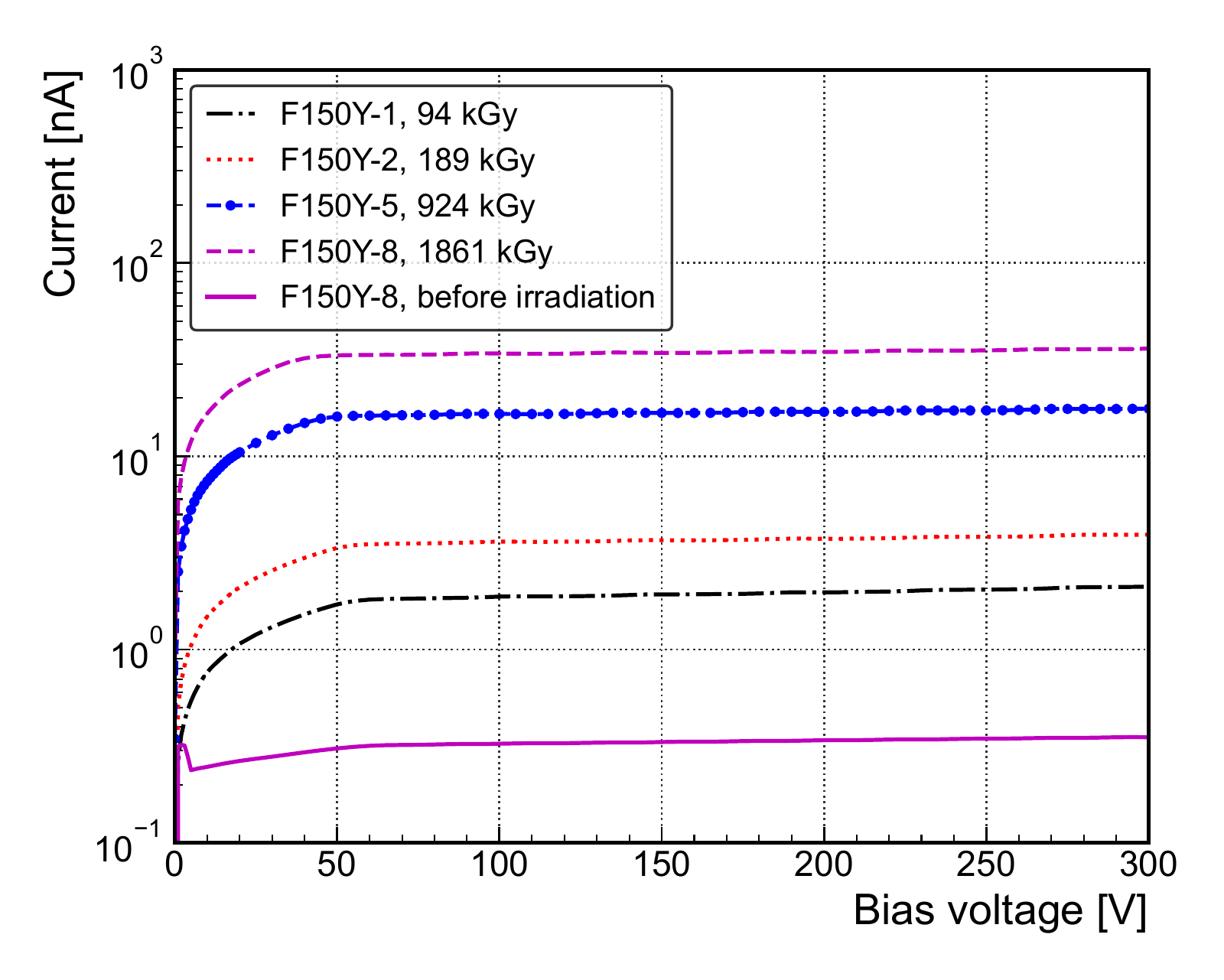}}\\
     \subfloat[\label{fig1b}]{
      \includegraphics[width=1.0\linewidth]{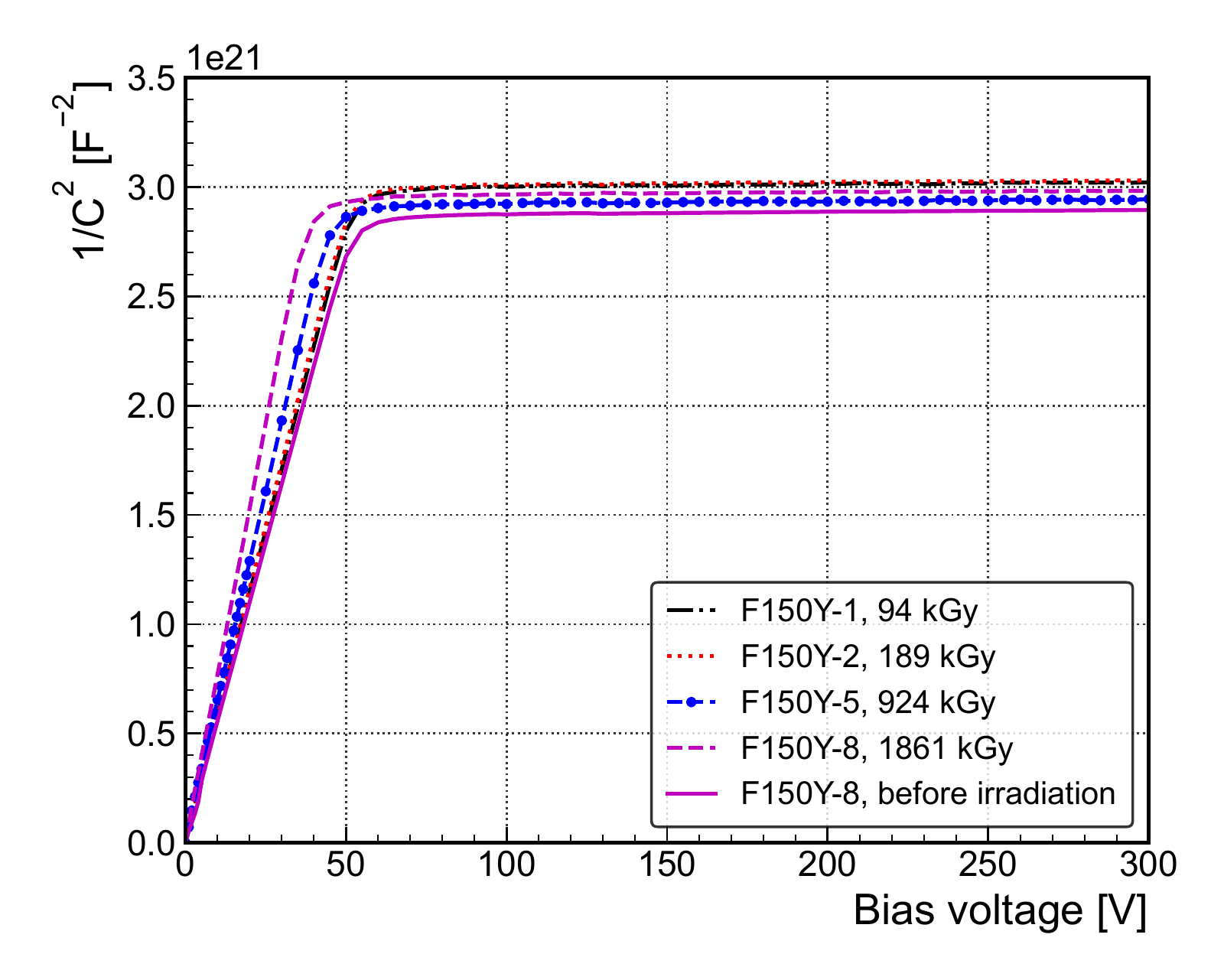}
    }
  \caption{(a) Current-voltage ($I$--$V$) characteristics of $p$-spray diodes irradiated to different dose values (see legend). (b) Capacitance-voltage ($1/C^2$-$V$) characteristics for the same diodes as in (a) for a frequency f = 500~kHz. Included in (a) and (b) are data for one diode before irradiation. Measurement condition: $T$ = \SI{20}{\celsius}, humidity $\leq$ \SI{10}{\percent}.}
  \label{fig1}
\end{figure}

\begin{figure}[!htb]
 \centering
  \subfloat[\label{fig2a}]{
    \includegraphics[width=1.0\linewidth]{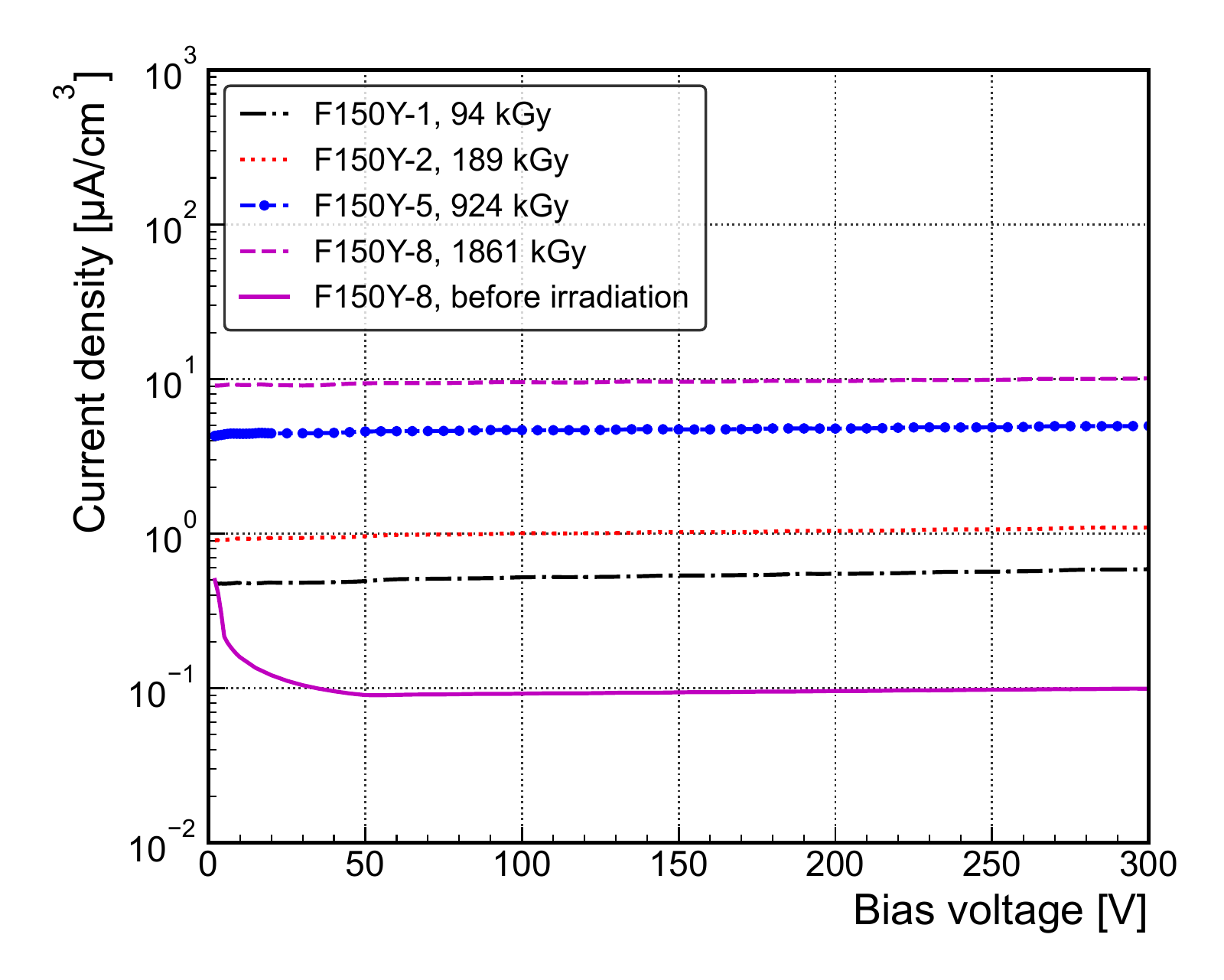}
   }\\
   \subfloat[\label{fig2b}]{ 
     \includegraphics[width=1.0\linewidth]{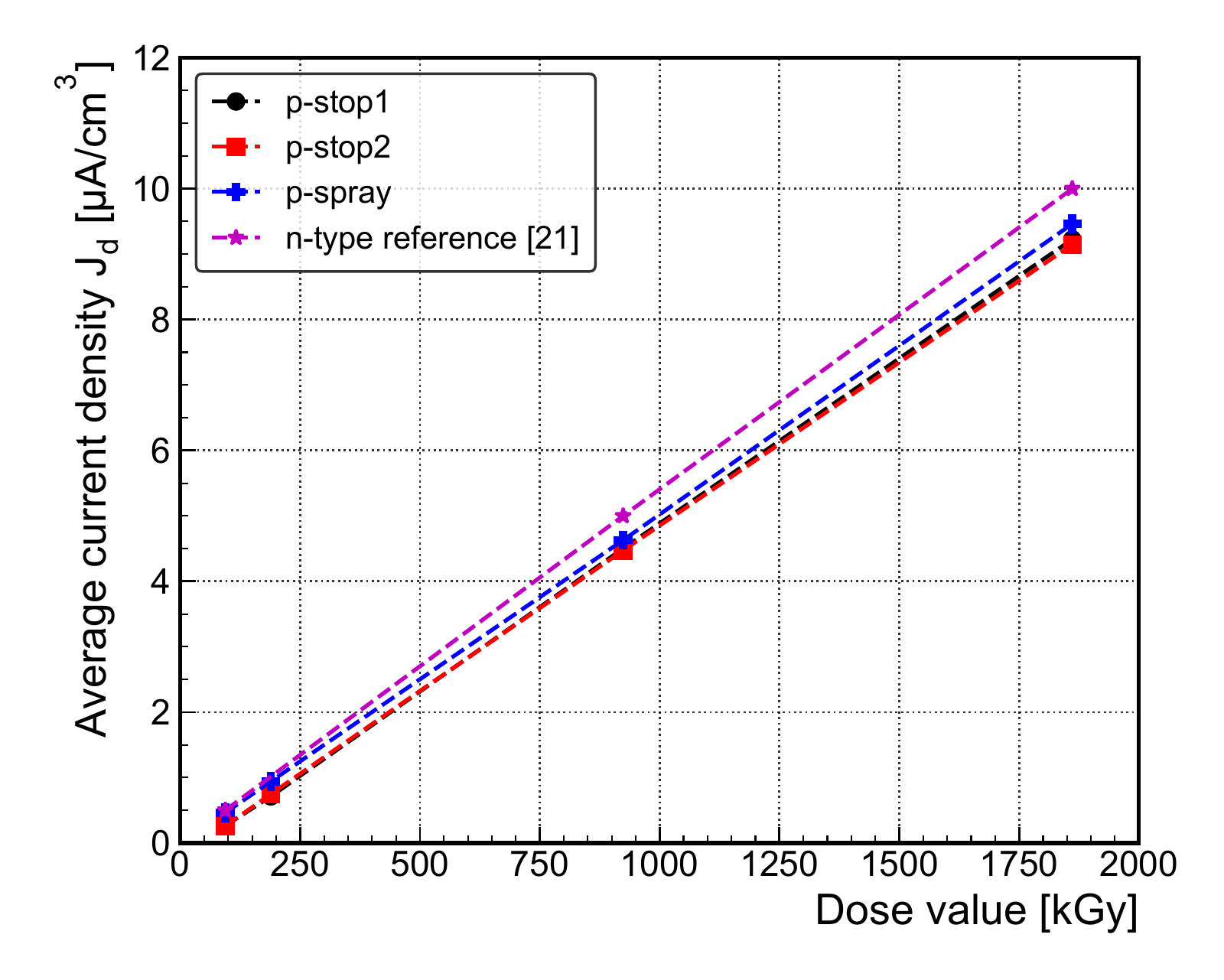}
  }
  \caption{(a) Density of leakage current ($j_d$) versus bias voltage $V$, which developed from Fig.~\ref{fig1a} and \ref{fig1b}. (b) Average current density $J_d$ vs. Dose value. The values were taken from (a) in the voltage range from 100 V to 150 V. (more details see text).}
\label{fig2}
\end{figure}
\par
In Fig.~\ref{fig1} are given the $I$--$V$ and the $1/C^2$--$V$ curves of all $p$-spray diodes irradiated to different dose values. Included are also the data for the F150Y-8 diode before being irradiated. The $N_\text{eff}$ depth profile was determined from $C$--$V$ measurements,considering the following relations for $N_\text{eff}(w(V))$ and the depletion depth $w(V)$:
\par
\begin{eqnarray}
N_\text{eff} (V) & = & \frac{2}{\epsilon_0 \epsilon_r A^2 q_0\,d(1/C^2)/dV} \label{eqn:1}\\
w (V) & = & \frac{\epsilon_0 \epsilon_r A}{C(V)} \label{eqn:2}
\end{eqnarray}
where $C$ is the measured capacitance, $\epsilon_0$ is the permittivity of vacuum, $\epsilon_r$ the relative permittivity of silicon (11.9), $q_0$ is the elementary charge, $A$ is the active pad area. Considering the voltage dependence of the depletion depth (Eq.(2)) the current density $j_d$($V$) can be calculated from the measured $I$-$V$ curves according to $j_d$($V$) = $I$($V$)/(A$\cdot$$w$($V$)). The results are shown in Fig.~\ref{fig2a}, for the different dose values applied to the corresponding $p$-spray diodes. Average current densities $J_d$ as taken from $j_d$($V$) data, in the 100~V and 150~V voltage range for $V_\text{bias}$, are plotted in Fig.~\ref{fig2b} as a function of the dose, for both, $p$-stop and $p$-spray diodes. The data achieved for $n$-type oxygen-enriched DOFZ diodes, taken from reference \cite{b20}, are also included. As can be observed in Fig.~\ref{fig2b}, all the presented results are nearly identical.

\par
For the current density $J_d$ of the $p$-spray type diodes, an error of 1\% was estimated. This includes the uncertainty of the measured current and the depleted volume extracted from the $C$--$V$ data. The corresponding value for the $p$-stop diodes was estimated to be 2\%. The definition of current-related damage parameters $\alpha$ can also be applied to $\gamma$-ray damage with dose value ($D$):
\par
\begin{equation}
  \alpha_\gamma = \frac{\Delta J_d}{\Delta D}
\end{equation}
\par
\begin{figure}[!htb]
\centering
\includegraphics[width=1.0\linewidth]{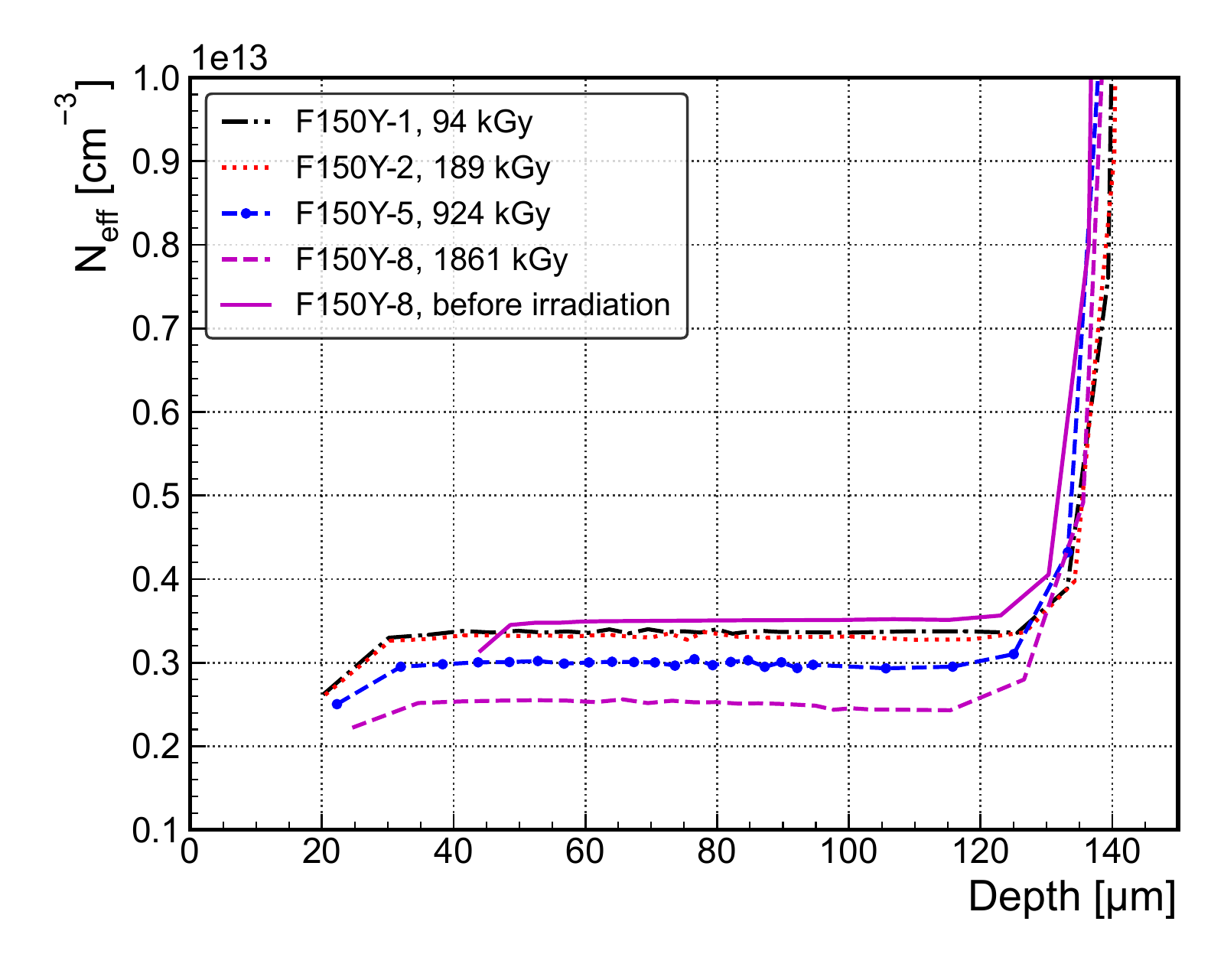}
\caption{$N_\text{eff}$ profile of the $p$-spray diodes for different dose values. The data were evaluated from $C$-$V$ measurements performed with a frequency of 500 kHz, by using Eq.~(\ref{eqn:1}) and Eq.~(\ref{eqn:2}).}
\label{fig3}
\end{figure}

According to Eq.(3) the $\alpha_\gamma$ value for the $p$-spray diodes is estimated to be $\alpha_\gamma$ = $\num{5.09e-12}$~$\si{A/(Gy \cdot cm^3)}$ with an error of 4.4\%. The value for the $p$-stop diodes is $\num{5.07e-12}$~$\si{A/(Gy \cdot cm^3)}$ with an error of 5.4\%. The errors are dominated by the uncertainties of the dose values (see Table~\ref{tab:table1}). Figure~\ref{fig3} presents the extracted $N_\text{eff}$ profiles for $p$-spray diodes, different irradiation dose values. With increasing the dose, the profiles are shifted to lower $N_\text{eff}$ values. Such a behaviour is expected due to the deactivation of the initial boron concentration caused by irradiation, the so-called acceptor removal effect \cite{b7, b8, b9, b10, b11}.

It should be mentioned here that the $C$--$V$ measurements were also performed for different frequencies (\SI{230}{\hertz}, \SI{455}{\hertz}, \SI{1}{\kilo\hertz}, \SI{10}{\kilo\hertz}, \SI{100}{\kilo\hertz}, \SI{200}{\kilo\hertz}, \SI{400}{\kilo\hertz}, \SI{500}{\kilo\hertz}) and a frequency dependence was observed only for $p$-stop diodes. This aspect will be discussed later in the subsection~\ref{sec:3.3} Surface effects.
\par

\begin{figure*}[!htp]
 \centering
  \subfloat[\label{fig4a}]{
   \includegraphics[width=0.49\linewidth]{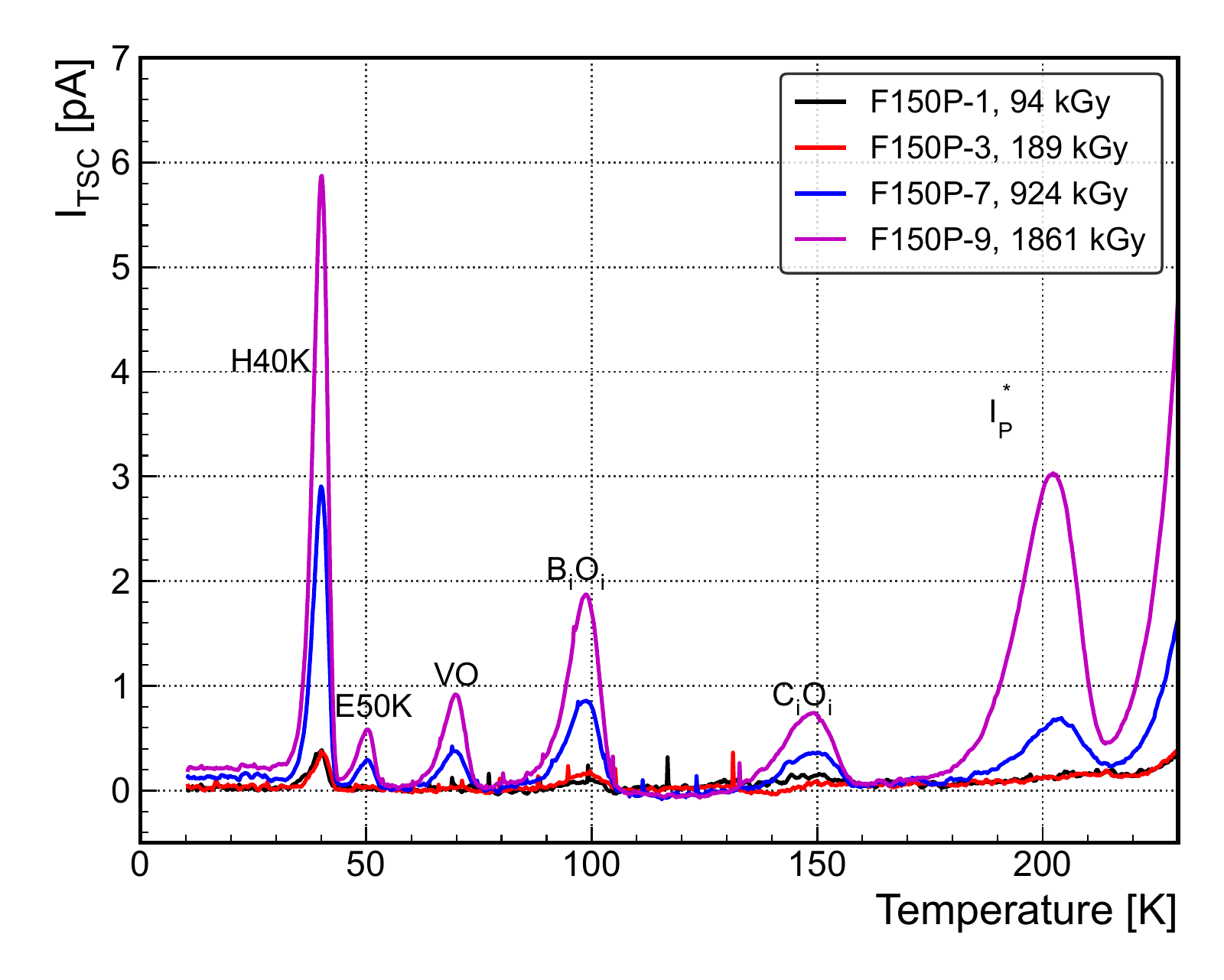}
   }
   \subfloat[\label{fig4b}]{
   \includegraphics[width=0.49\linewidth]{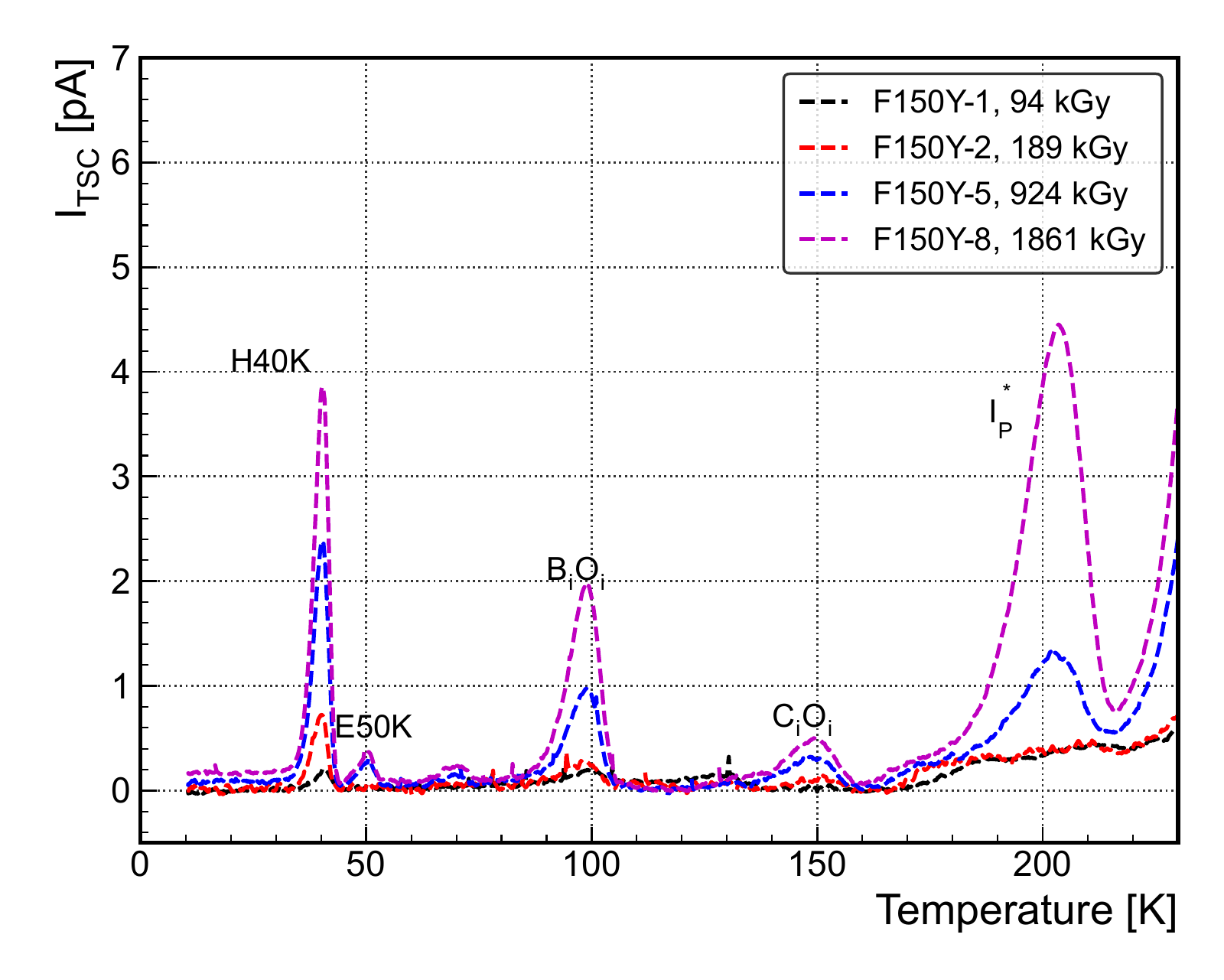}
   } \\
   \subfloat[\label{fig4c}]{ 
    \includegraphics[width=0.49\linewidth]{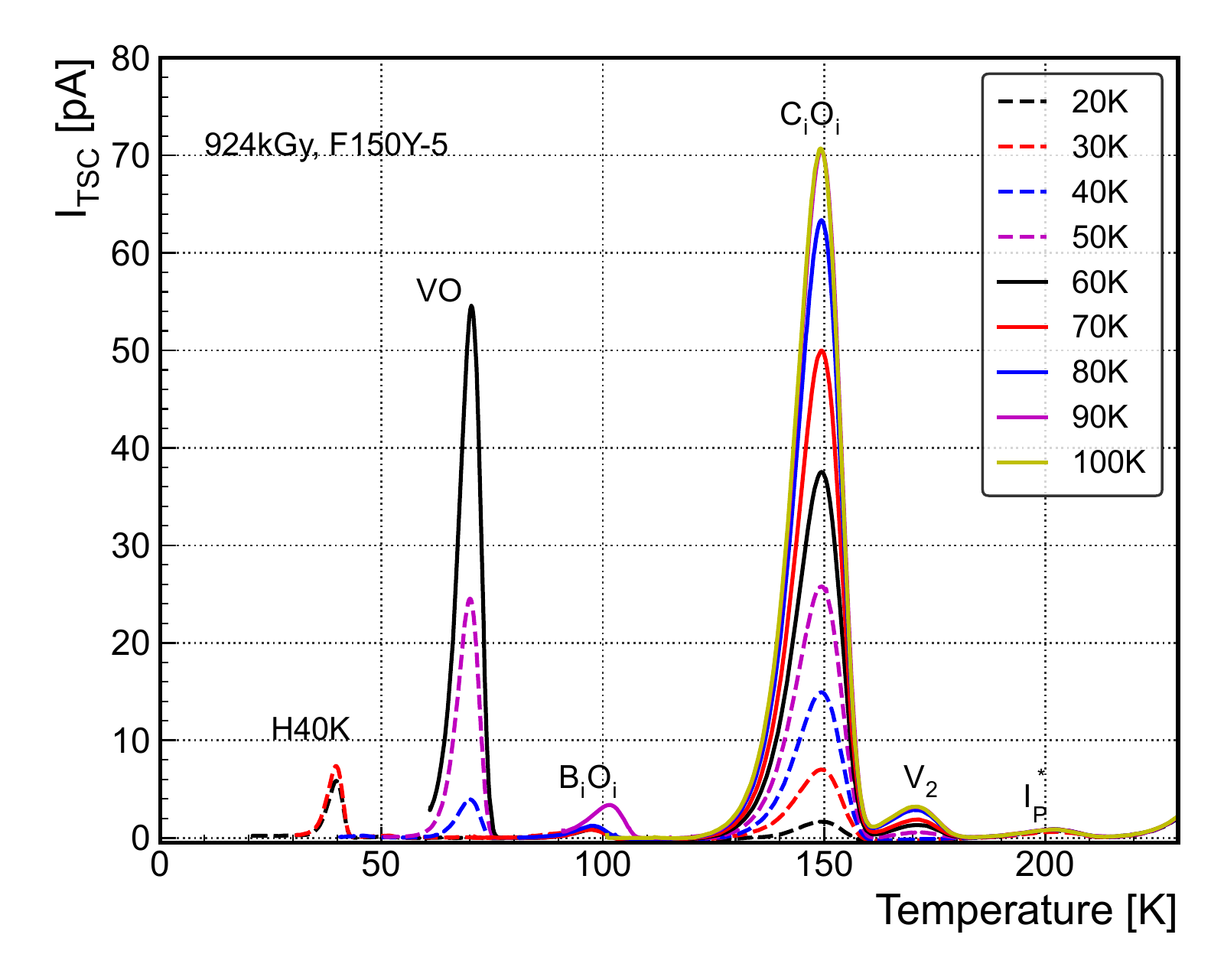}
    }
   \subfloat[\label{fig4d}]{
      \includegraphics[width=0.49\linewidth]{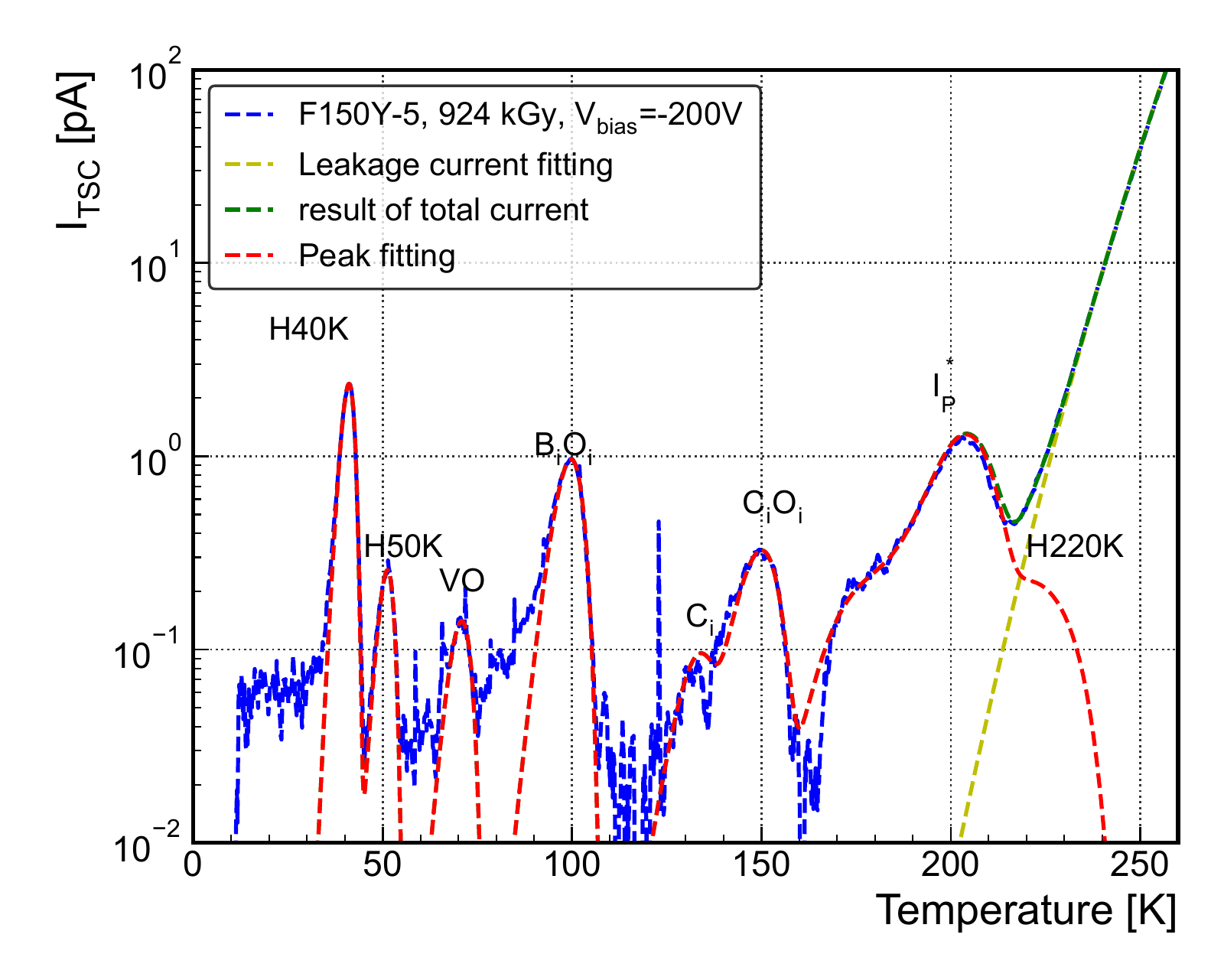}
      } 
\caption{TSC spectra: (a) for $p$-stop diodes and (b) for $p$-spray diodes. In both cases, the spectra correspond to a reverse bias of $V_{bias}$ = -200~V and were recorded after filling the traps by injecting a 1~mA forward current ($V_{fill}$ = 10~V) at 10~K. (c) after the same filling procedure as in a) and b), but with different $T_{fill}$ for diode F150Y-5. (d) experimental and computed by fitting the different peaks for diode F150Y-5.}
\label{fig4}
\end{figure*}

In order to investigate the radiation-induced defect complexes by $^\text{60}$Co $\gamma$-rays in the high resistivity FZ diodes, the TSC method is used. Figure~\ref{fig4a} and \ref{fig4b} show TSC spectra of $p$-stop and $p$-spray diodes irradiated to different dose values, respectively. All spectra in Fig.~\ref{fig4} were measured on fully depleted sensors ($V_\text{bias} = \num{-200}~\si{V}$, $V_\text{dep} = \num{-50}~\si{V}$) over the entire TSC temperature range. As can be seen in Fig.~\ref{fig4}, several peaks are induced after irradiation and increase with dose. Some of these traps corresponded to the previously detected radiation induced defect complexes H40K~\cite{b5}, VO~\cite{b21}, B$_\text{i}$O$_\text{i}$~\cite{b3}, C$_\text{i}$O$_\text{i}$~\cite{b13, b21} and V$_2$~\cite{b6}. A TSC peak, similar to the one labelled as I$_\text{P}^*$ in Fig.~\ref{fig4}, has been reported as I$_\text{P}$ defect in previous studies and associated with V$_2$O complex, a defect predominantly generated via a second order process in oxygen-lean material, thermally stable up to \SI{350}{\celsius} and with a strong impact on both, the leakage current and $N_\text{eff}$~\cite{b22, b23, b24, b25}. However, the trapping parameters determined for the I$_\text{P}^*$ defect are different compared with of I$_\text{P}$ defect - see Table~\ref{tab:table2}. Accordingly, also the impact of the I$_\text{P}^*$ defect on both leakage current and $N_\text{eff}$ are negligible. By comparing Fig.\ref{fig4a} and Fig.\ref{fig4b}, it can be observed that some of the peaks observed in the $p$-spray diodes are smaller compared to those recorded in the $p$-stop ones (H40K, E50K, VO and C$_\text{i}$O$_\text{i}$). This might be due to the different filling of the defects at 10 K. As demonstrated in Fig.~\ref{fig4c} most peaks show a strong dependence on the filling temperature, except the B$_\text{i}$O$_\text{i}$ and I$_\text{P}^*$ traps. Previous studies have shown that for the C$_\text{i}$O$_\text{i}$ defect an extremely strong dependence on the filling temperature ($T_\text{fill}$) exists and thus, suggesting a multi-phonon capture process~\cite{b13, b26}. To get the full defect concentration TSC measurements with different $T_\text{fill}$ (from 20$\sim$100~K in steps of 10~K) were performed.

\par
As can be seen in Fig.~\ref{fig4c}, an increase of the peak height with increasing $T_\text{fill}$ is not only observed for the C$_\text{i}$O$_\text{i}$ defect but also for the H40K, VO and the V$_2$ defect complexes. Only the B$_\text{i}$O$_\text{i}$ and I$_\text{P}^*$ have the same peak maximum for different $T_\text{fill}$. The abrupt increase in the B$_\text{i}$O$_\text{i}$ peak after filling at 90~K is attributed to the higher and not constant heating rate at the initial stage of the temperature ramp. In the fully depleted diodes, the general formula of thermally stimulated current $I^{e}_{TSC}(T)$ or $I^{h}_{TSC}(T)$ for a single electron or hole trap is given by \cite{b13, b17, b18, b27, b28}:
\par
\begin{eqnarray}
I^{e}_{TSC} (T) & = & { {1 \over 2} \cdot q_0 \cdot A \cdot d \cdot e_{n}(T)
\cdot n_t(T_{fill}) \cdot f\left(T\right)} \\
I^{h}_{TSC} (T) & = & { {1 \over 2} \cdot q_0 \cdot A \cdot d \cdot e_{h}(T)
\cdot p_t(T_{fill}) \cdot f\left(T\right)} \\
e_{n,p} & = & {\sigma_{n,p} \cdot v_{th,n,p} \cdot N_{C,V} \cdot \exp{\left(-{E_a \over
{k_B T}}\right)}}\\
f(T) & = & {\exp{\left(- {1 \over{\beta}}
\int_{T_0}^{T}{\left(e_n(T')+e_p(T')\right)} \, dT'\right)}}\
\end{eqnarray}
where $T$ is the measured temperature, $d$ is the fully depleted depth. $e_n$ and $e_p$ are the emission rates for electrons and holes. $v_{th,n,p}$ and $N_{C,V}$are the thermal velocity of electrons/holes and the density of states in the conduction/valence band, respectively, both taken from~\cite{b29}. The activation energy is defined as $E_a$ = $E_C - E_t$ or $E_a$ = $E_t - E_V$ according to the type of emitted charge, electrons or holes, respectively, with $E_t$ being the energy level of the trap and $E_{C,V}$ the conduction/valence band edges. The $\sigma_{n,p}$ terms stand for the capture cross-section of electrons/holes, $k_B$ is the Boltzmann constant and the $f(T)$ function is the defect occupancy at temperature $T$. The $n_t(T_{fill})$ and $p_t(T_{fill})$ terms represent the density of defects that are filled with electrons or holes after injection at temperature $T_\text{fill}$. The considered trapping parameters are given in Table~\ref{tab:table2} and kept fixed in the peak fitting procedure.

\begin{table*}[htbp]
\centering
\caption{The trapping parameters of the detected defects. An example of how much these defects contribute to the change in $N_\text{eff}$ and $J_d$ at RT is given for F150Y-5.}
  \begin{tabular}{@{}lccccccccc@{}}
   \toprule
   label & H40K & E50K & VO & B$_\text{i}$O$_\text{i}$ & C$_\text{i}$ & C$_\text{i}$O$_\text{i}$ & V$_\text{2}$ & I$_\text{P}^*$ & H220K \\
   Reference & \cite{b13} & \cite{b13} & \cite{b23} & \cite{b11} & \cite{b13} & \cite{b21} & \cite{b13} & This work & \cite{b13} \\
  \midrule
  $\sigma_p$ (10$^{-15}$ $\si{cm^2}$) & 4.3 & - & - & $\num{1e-5}$ & 4.28 & 0.94 & - & 0.01 & 29.2\\  
  $\sigma_n$ (10$^{-15}$ $\si{cm^2}$) & - & 5.4 & 6.1 & 10.5 & - & - & 1.5 & 1.45 & 0.6 \\
  $E_a$ (eV) & 0.108 & 0.110 &  0.160 & 0.258 & 0.317 &  0.367 & 0.429 & 0.523 & 0.564 \\
    & ($E_t$-$E_V$) & ($E_C$-$E_t$) & ($E_C$-$E_t$) & ($E_C$-$E_t$) & ($E_t$-$E_V$) & ($E_t$-$E_V$) & ($E_C$-$E_t$) & ($E_C$-$E_t$) & ($E_t$-$E_V$) \\
  $N_\text{eff}$ at RT ($\si{cm^{-3}}$) & 0 & 0 & 0 & -$\num{2.5e11}$ & 0 & -$\num{4e4}$ & $\num{1.9e8}$ & $\num{1.6e8}$ & -$\num{1e11}$ \\
  $J_d$ at RT (A/$\si{cm^3}$) & 0 & 0 & 0 & 0 & 0 & $\num{3e-11}$ & $\num{1.9e-6}$ & $\num{2.2e-8}$ & $\num{1.3e-6}$ \\
  $N_t$ of F150Y-5 ($10^{12}~\si{cm^{-3}}$) & 0.53 & 0.02 & 5.6 & 0.25 & 0.02 & 17 & 0.91 & 0.31 & 0.11 \\
  \bottomrule
  \end{tabular}
\label{tab:table2}
\end{table*}

\par
For describing the leakage current, $I_{LC}$ measured at temperatures $T$ $\geq$ \SI{230}{\kelvin}, the following equation is used \cite{b30}:
\begin{eqnarray} 
I_{LC} & = & { \chi \cdot {T^2} \cdot \exp{\left(-{{\delta E} \over {k_B T}}\right)}} 
\end{eqnarray}
where the $\chi$ and $\delta$E are free parameters in the fitting procedure. The extracted values are achieved from fits to the data in the temperature ranges from 230~K to 250~K at $V_\text{bias}$ = -200~V, which are included in Table~\ref{tab:table3}. For both parameters, the errors were estimated to be below 1\%.

\begin{table*}[htbp]
\centering
\caption{Extracted fit parameters for the leakage current LC ($V_\text{bias}$ = -200~V) by using Eq.~8}
  \begin{tabular}{@{}lccccc@{}}
   \toprule
    & F150P-1 (94 kGy) & F150P-3 (189 kGy) & F150P-7 (924 kGy) & F150P-8 (1861 kGy) \\
  \midrule
  $p$-stop $\chi$ ($\si{A/K}$) &  $\num{0.0021}$ & $\num{0.0204}$ & $\num{0.1273}$ & $\num{0.4415}$\\
  $p$-stop $\delta E$ ($\si{eV}$) &  $\num{0.6594}$ & $\num{0.7103}$ & $\num{0.7103}$ & $\num{0.7196}$\\
  \midrule 
   & F150Y-1 (94 kGy) & F150Y-2 (189 kGy) & F150Y-5 (924 kGy) & F150Y-8 (1861 kGy) \\
  \midrule 
  $p$-spray $\chi$ ($\si{A/K}$) &  $\num{0.0005}$ & $\num{0.0037}$ & $\num{0.1260}$ & $\num{0.2002}$\\
  $p$-spray $\delta E$ ($\si{eV}$) &  $\num{0.6410}$ & $\num{0.6744}$ & 
   $\num{0.7103}$ & $\num{0.7103}$\\
  \bottomrule
  \end{tabular}
\label{tab:table3}
\end{table*}

\par

\begin{figure*}[!htp]
 \centering
  \subfloat[\label{fig5a}]{
   \includegraphics[width=0.49\linewidth]{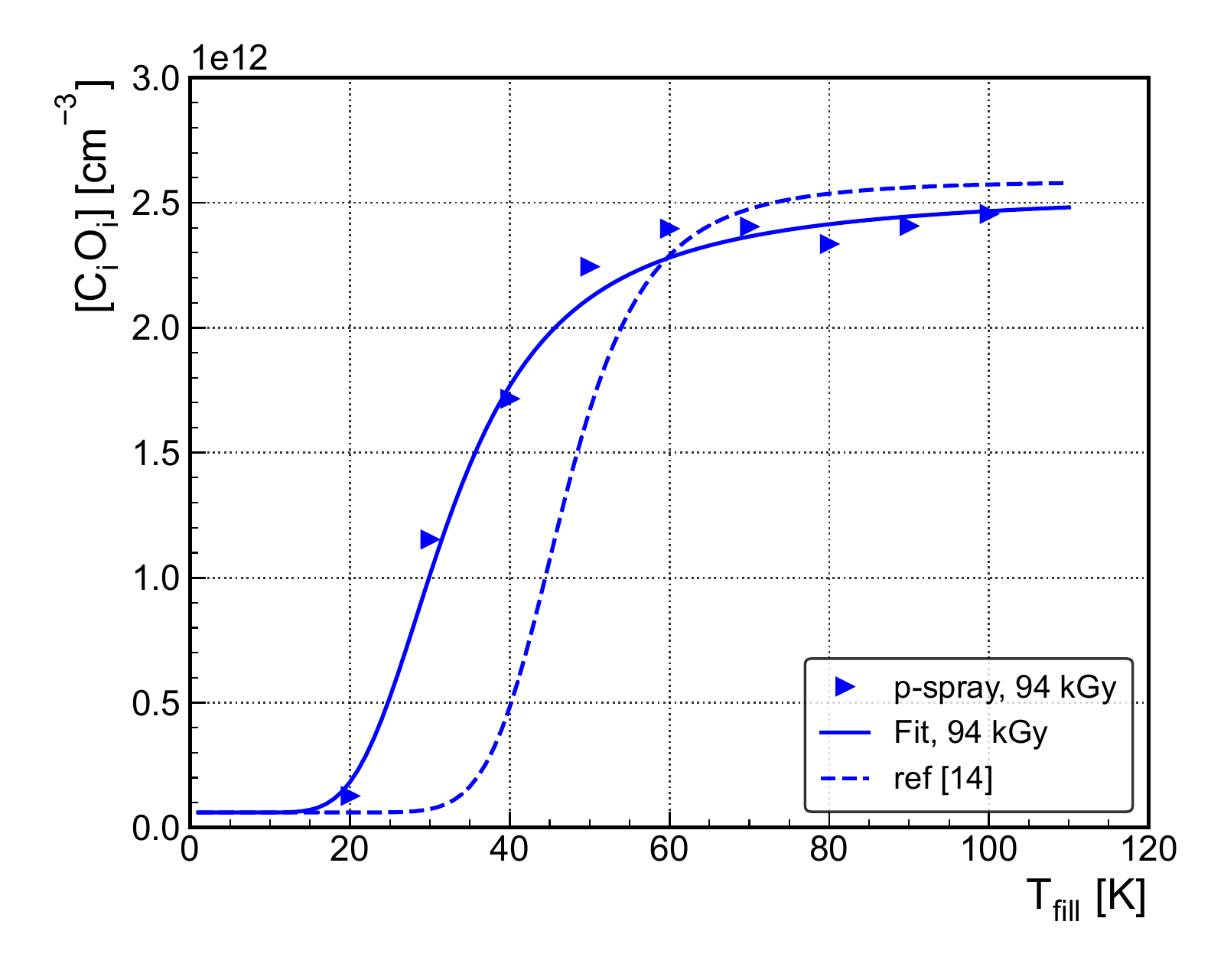}
   }
   \subfloat[\label{fig5b}]{
   \includegraphics[width=0.49\linewidth]{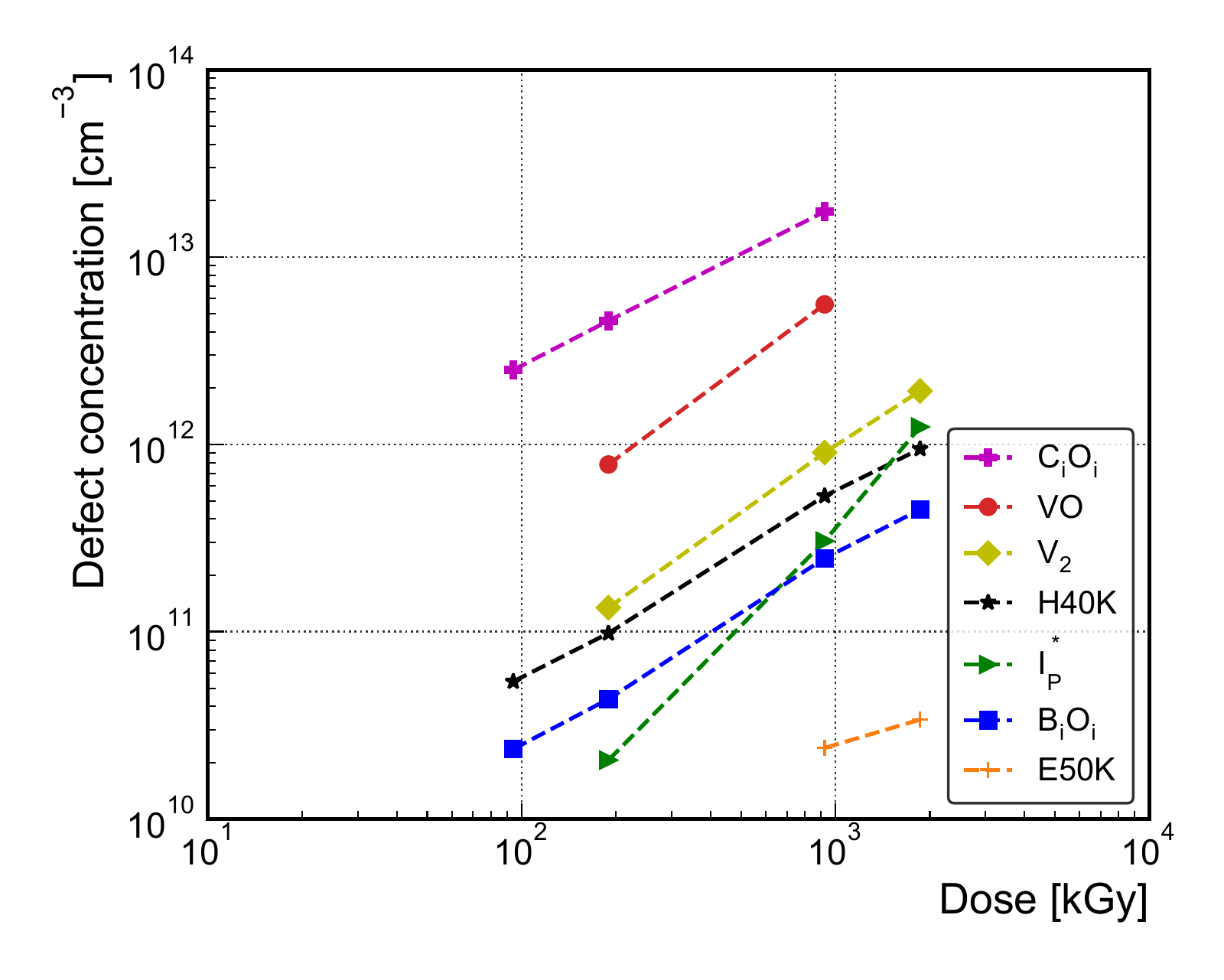}
      } 
\caption{(a) [C$_\text{i}$O$_\text{i}$] vs $T_\text{fill}$, which included the data, our fits and literature value. The data of [C$_\text{i}$O$_\text{i}$] are extracted from TSC spectra e.g.\ Fig.~\ref{fig4}c. The fit results are given by fitting the TSC spectra using Eq.~8.
(b) Defect concentration vs.\ Dose value included all observed defects.}
\label{fig5}
\end{figure*}

The concentrations of defects were determined by fitting the measured TSC peaks using the equations Eq.~(4-7). An example of fitting a TSC spectrum is presented in Fig.~\ref{fig4d} for a $p$-spray diode irradiated with a dose of 94~kGy. The dependence of the C$_\text{i}$O$_\text{i}$ concentration ([C$_\text{i}$O$_\text{i}$]) on the filling temperature $T_\text{fill}$ is shown in Fig.~\ref{fig5a}. As it can be observed, [C$_\text{i}$O$_\text{i}$] initially increases with $T_\text{fill}$ and saturates at $T_{fill} \geq$ 60~K. To describe the $T_\text{fill}$ dependence of [C$_\text{i}$O$_\text{i}$] the following equation is used~\cite{b13, b26, b31}:
\begin{eqnarray} 
n_t (T_{fill}) & = & N_{offset} + { N_t \times {1 \over{1+a \cdot \exp \left({E_s \over{k_B T_{fill}} }\right) } }} 
\end{eqnarray}
The offset value is caused by the background current. $N_t$ is the total C$_\text{i}$O$_\text{i}$ concentration [C$_\text{i}$O$_\text{i}$], and $a$ is a constant. $E_s$ is the activation energy for non-radiative multi-phonon capture of a charge carrier, accounting for the temperature dependence of the capture cross sections~\cite{b31}. The term $a \cdot \exp \left({E_s \over{k_B T_{fill}} }\right)$ is given by the ratio ${n \cdot c_n} \over {p \cdot {c_p}}$, where $n$ and $p$ are the concentrations of free electrons and holes under forward bias injection and $c_{n,p}$ are the capture coefficients of the corresponding charge carriers.
\par
The parameters obtained for fitting the C$_\text{i}$O$_\text{i}$ data in Fig.~\ref{fig5a} are: $a$ = (1.18 $\pm$ 0.03)~$\times 10^{-2}$, $E_s$ = (13.2 $\pm$ 0.2)~meV, $N_{offset}$ = (6.1 $\pm$ 0.5)~$\times 10^{10}$~$\text{cm}^{-3}$, and $N_t$ = (2.5 $\pm$ 0.1)~$\times 10^{12}$~$\text{cm}^{-3}$. The fitted result is plotted in Fig.~\ref{fig5a} together with the experimental data. We also included a calculated curve using the values of $a$ = $10^{-4}$ and $E_s$ = 37.3~meV given in a previous study on neutron irradiated diodes~\cite{b13}, considering the same values of $N_{offset}$ and $N_t$ as in our data fit. The different $a$ and $E_s$ values in our fit compared with those from Moll~\cite{b13} might be explained by the potential barriers surrounding the clustered regions induced by neutron irradiation, which can slow down the carriers capturing process~\cite{b41, b42}.

\par

The increase with dose in the concentration of the TSC detected defects on the p-spray diode is shown in Fig.~\ref{fig5b}. Worth noting is that among all the defects, only the I$_\text{P}*$ defect does not have a linear dependence on Dose, but a quadratic one (slope $\sim$ 2 in Fig.~\ref{fig5b}) as previously reported for the I$_\text{P}$ defect, a radiation-induced defect formed via a second order process. Considering the definition of the introduction rate $g$~=~$\Delta N_t \over {\Delta D}$, the rates for $\gamma$-ray irradiation were extracted from the linear fits to the data (see Fig.~\ref{fig5b}) and are summarized in Table~\ref{tab:table4}.

\begin{table*}[htbp]
\centering
\caption{Introduction rates}
  \begin{tabular}{@{}lccccccc@{}}
   \toprule
    Defect & H40K & E50K & VO & B$_\text{i}$O$_\text{i}$ & C$_\text{i}$O$_\text{i}$ & V$_\text{2}$   \\
  \midrule
  Introduction rate ($\si{cm^{-3} \cdot Gy^{-1}}$) & $\num{5.2e5}$ & $\num{1.1e4}$ & $\num{6.5e6}$ &$\num{2.5e5}$ & $\num{1.9e7}$ & $\num{1.0e6}$ \\ 
  \bottomrule
  \end{tabular}
\label{tab:table4}
\end{table*} 

\par

It should be mentioned here that the errors in the extracted defect concentrations are caused by different reasons:
\begin{enumerate}
 \item[a)] The fitting procedure of the peak maxima leads to about 2\% error in the determined defect concentration.
 \item[b)] Errors of about 3\% are estimated for calculating the $n_t(T_\text{fill})/N_t$ fraction of filled defects at low temperature by forward current injection, due to the filling temperature dependence of in the case of H40K, C$_\text{i}$O$_\text{i}$ and V$_\text{2}$.
 \item[c)] The noise of the TSC signal contributes to an error in the determined defect concentration of about 2\%. 
\end{enumerate}
Assuming all errors to be uncorrelated, the total error on the defect concentration is below 5\%.
\par

\subsection{Annealing studies}\label{sec:3.2}
\begin{figure*}[!htp]
 \centering
  \subfloat[\label{fig6a}]{
   \includegraphics[width=0.49\linewidth]{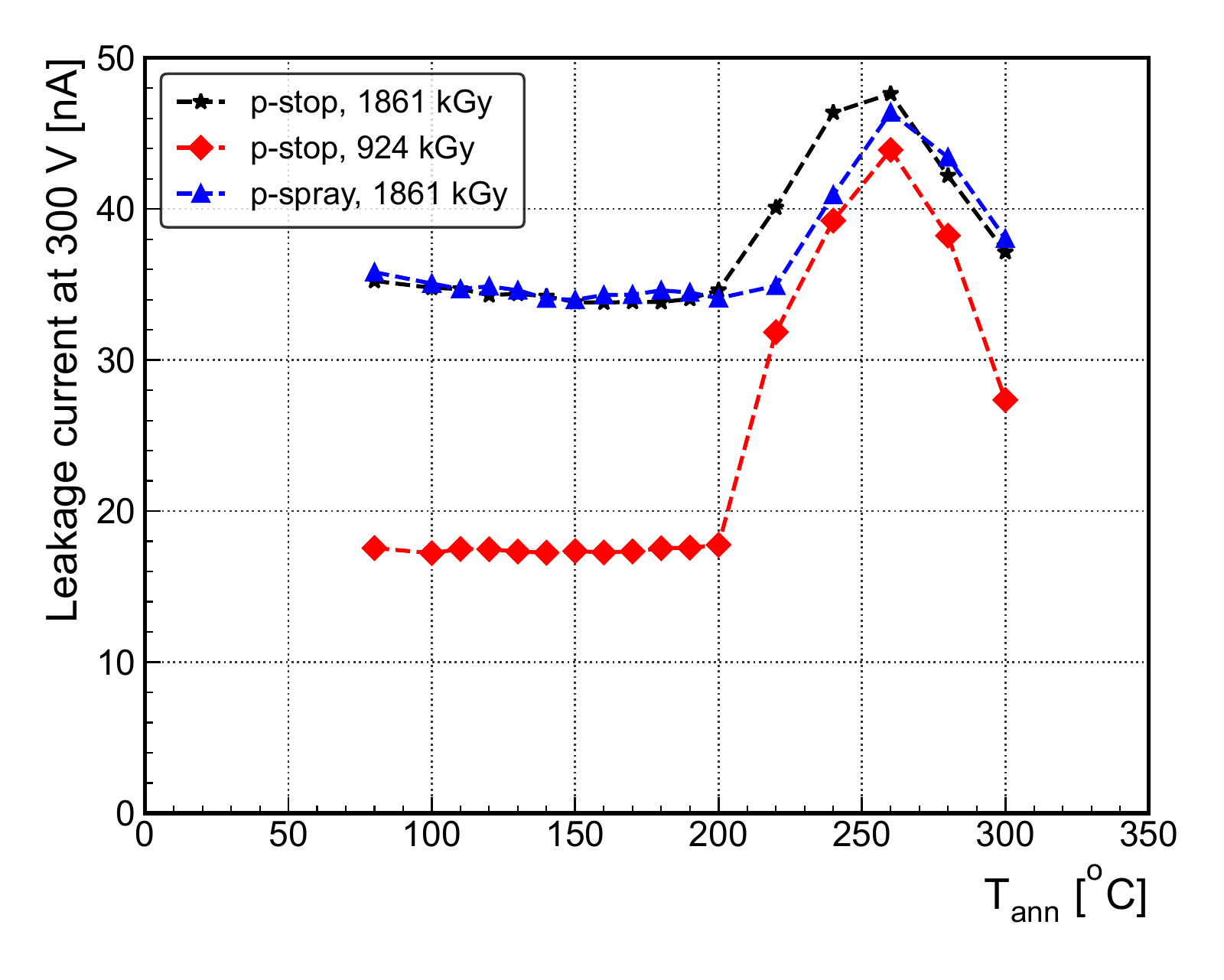}
   }
   \subfloat[\label{fig6b}]{
   \includegraphics[width=0.49\linewidth]{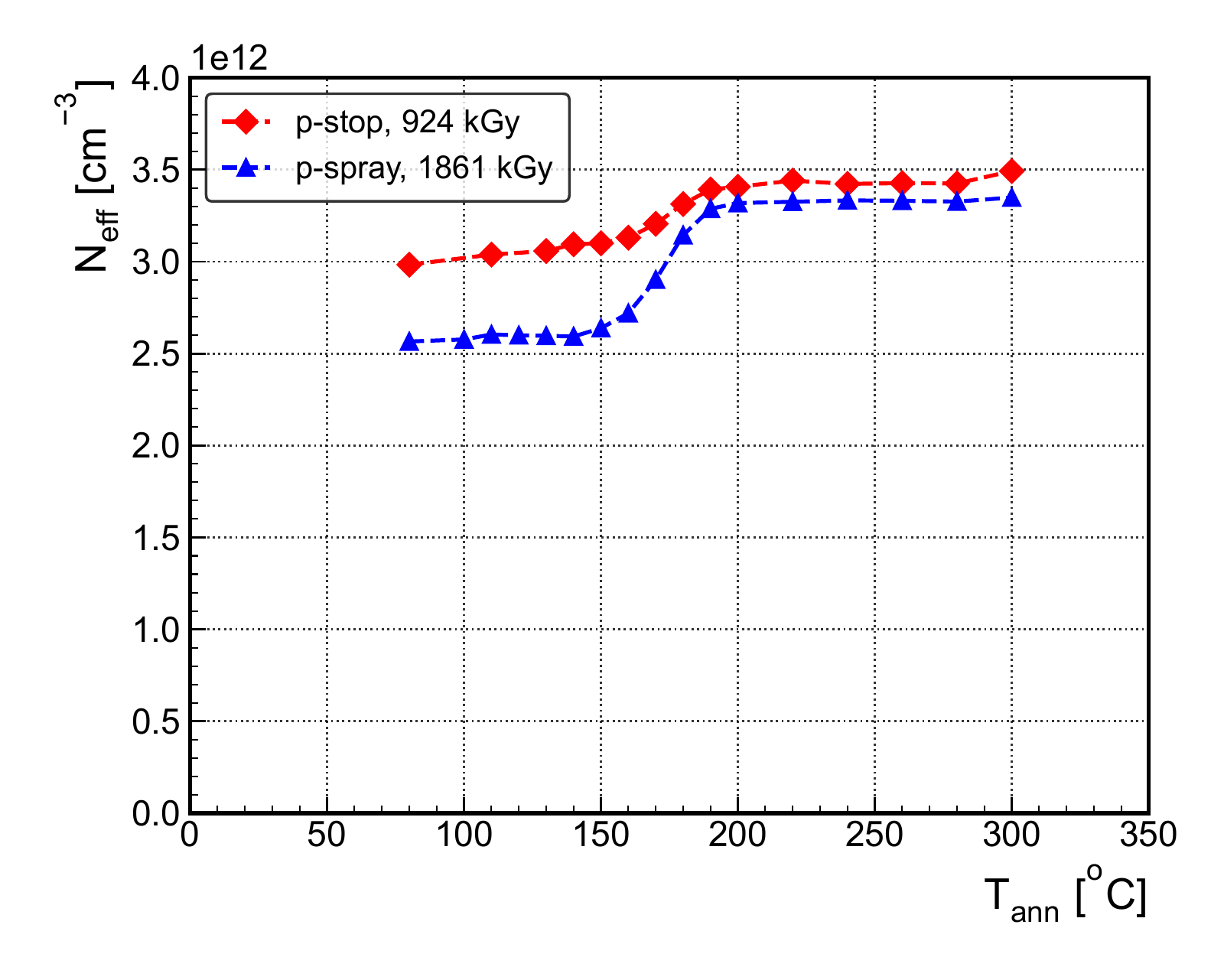}
        }
\caption{(a) The development of leakage current measured at room temperature and $V$ = -300~V with annealing temperature. (d) Extracted $N_\text{eff}$ from $C$--$V$ measurement at room temperature as a function of annealing temperature.}
\label{fig6}
\end{figure*}

Two p-stop diodes, F150P-7 and F150P-8, and one p-spray diode F150Y-8 have been subjected to annealing experiments and the changes in  the macroscopic and microscopic properties of these diodes have been studied. Isochronal annealing experiments have been performed for 15 min at different temperatures, between \SI{100}{\celsius} and \SI{300}{\celsius}. The temperature was increased in steps of \SI{10}{\celsius} for annealing up to \SI{200}{\celsius} and in steps of \SI{20}{\celsius} in the higher temperature range. The annealing behaviour of the reverse current at -300 V is plotted in Fig.~\ref{fig6a} for the three annealed irradiated diodes. As can be seen, while the change of the leakage current for $T_\text{ann}$ $\leq$ \SI{200}{\celsius} is very small, it becomes significant for higher temperatures for all the diodes. A sudden increase of the leakage current takes place in the \SI{200}{\celsius}$\sim$\SI{260}{\celsius} temperature range, being followed by a sharp decrease for higher temperatures. The reason for this sudden change is not clear yet.

\par
The development of $N_{\text{eff}}$ with annealing temperature is presented in Fig.~\ref{fig6b}. The given $N_{\text{eff}}$ data represents the average values determined from the $N_\text{eff} (w)$ profiles in the \SI{60}{\micro \meter} and \SI{100}{\micro \meter} depth range. Between \SI{150}{\celsius} and \SI{200}{\celsius} the $N_{\text{eff}}$ increases, recovering most of the initial doping value of not-irradiated devices $\num{3.5e12}$~$\si{cm^{-3}}$, e.g.\ 98\% for F150P-7 and 95\% for F150Y-8. This recovery of the acceptor doping is related to the dissociation of the B$_\text{i}$O$_\text{i}$ defect which anneals out in the \SI{150}{\celsius}-\SI{200}{\celsius} temperature range~\cite{b3, b33, b43, b44}.

\par

\begin{figure*}[!htp]
 \centering
  \subfloat[\label{fig7a}]{
   \includegraphics[width=0.49\linewidth]{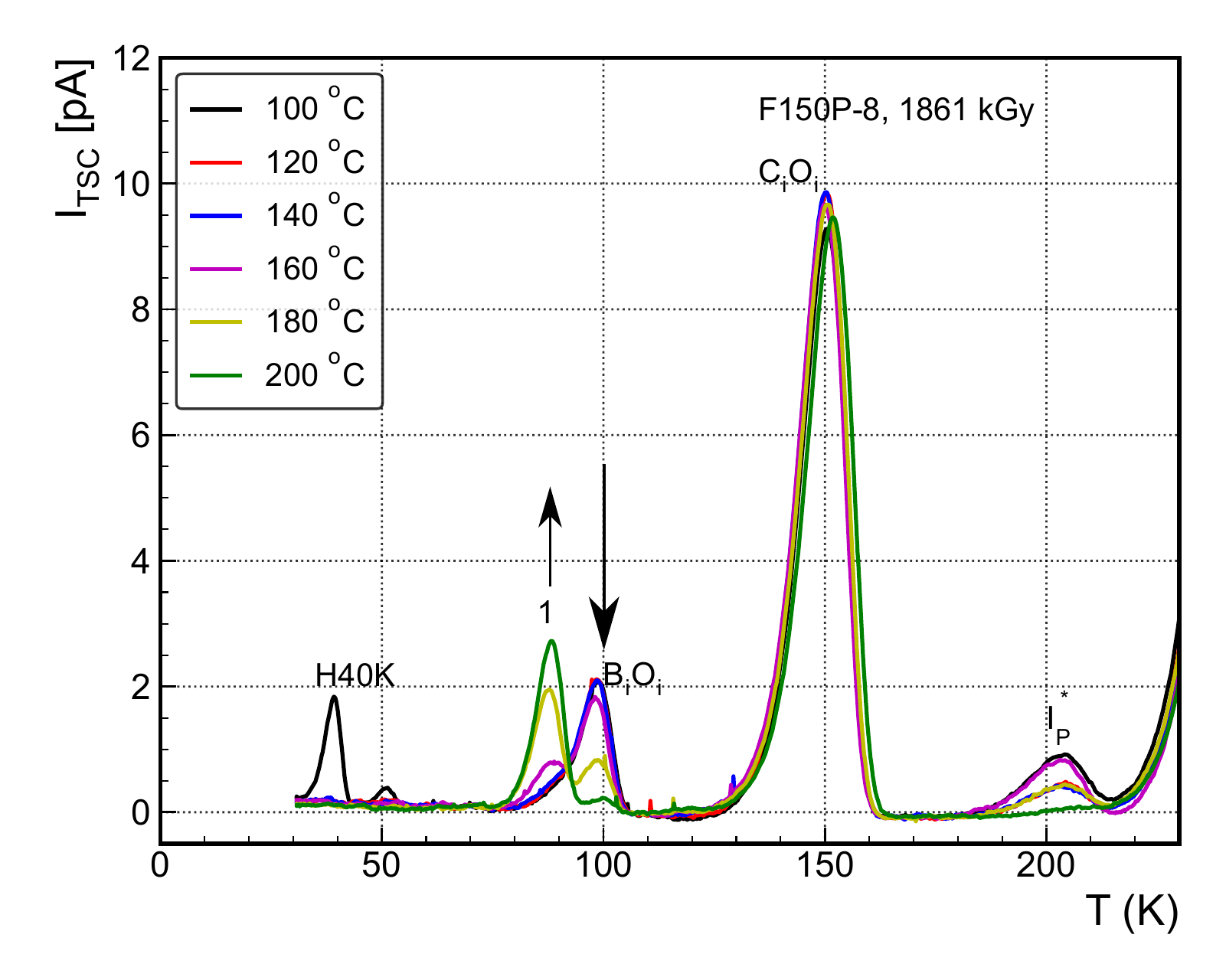}
   }
   \subfloat[\label{fig7b}]{
   \includegraphics[width=0.49\linewidth]{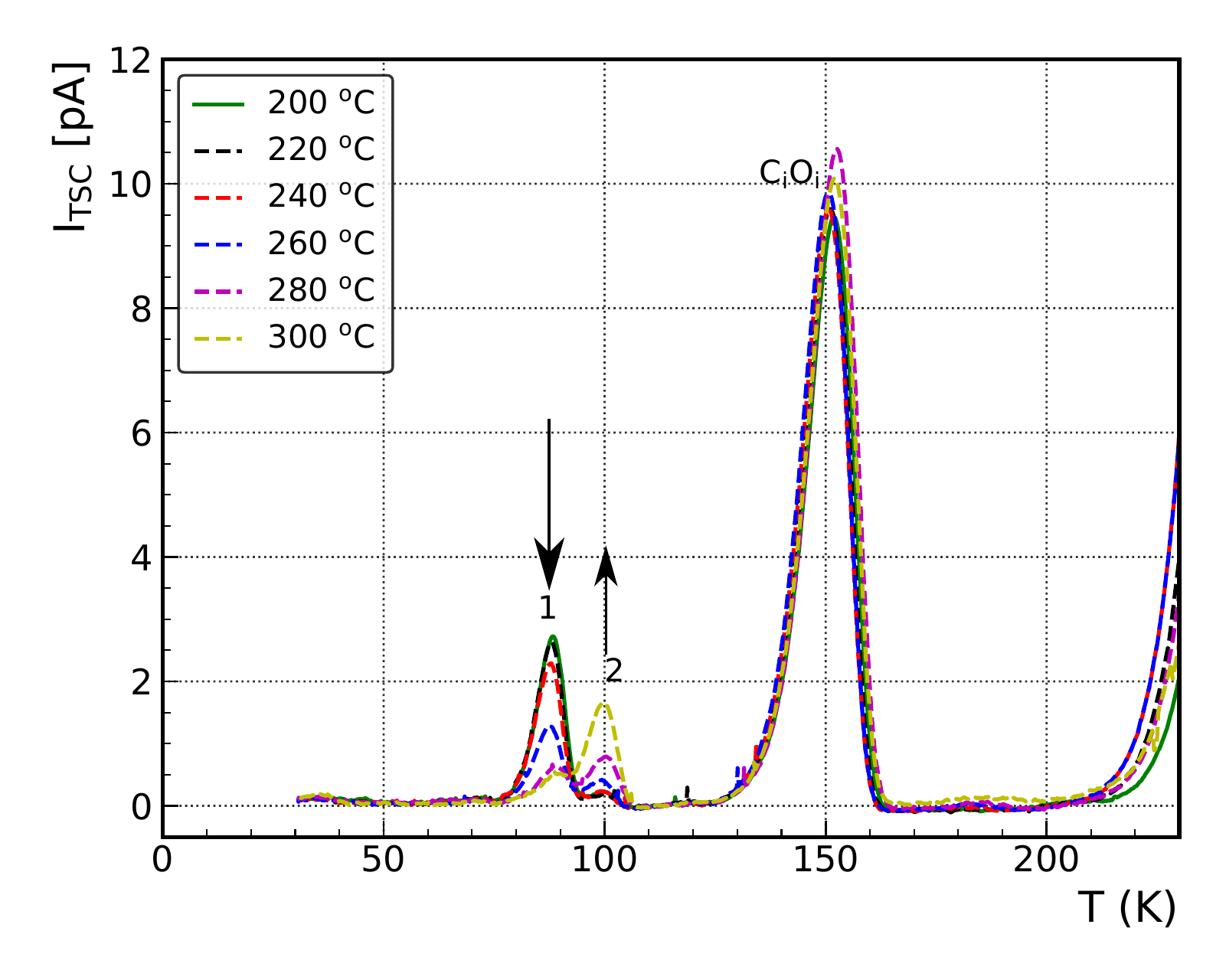}
      }
      \\
   \subfloat[\label{fig7c}]{
   \includegraphics[width=0.49\linewidth]{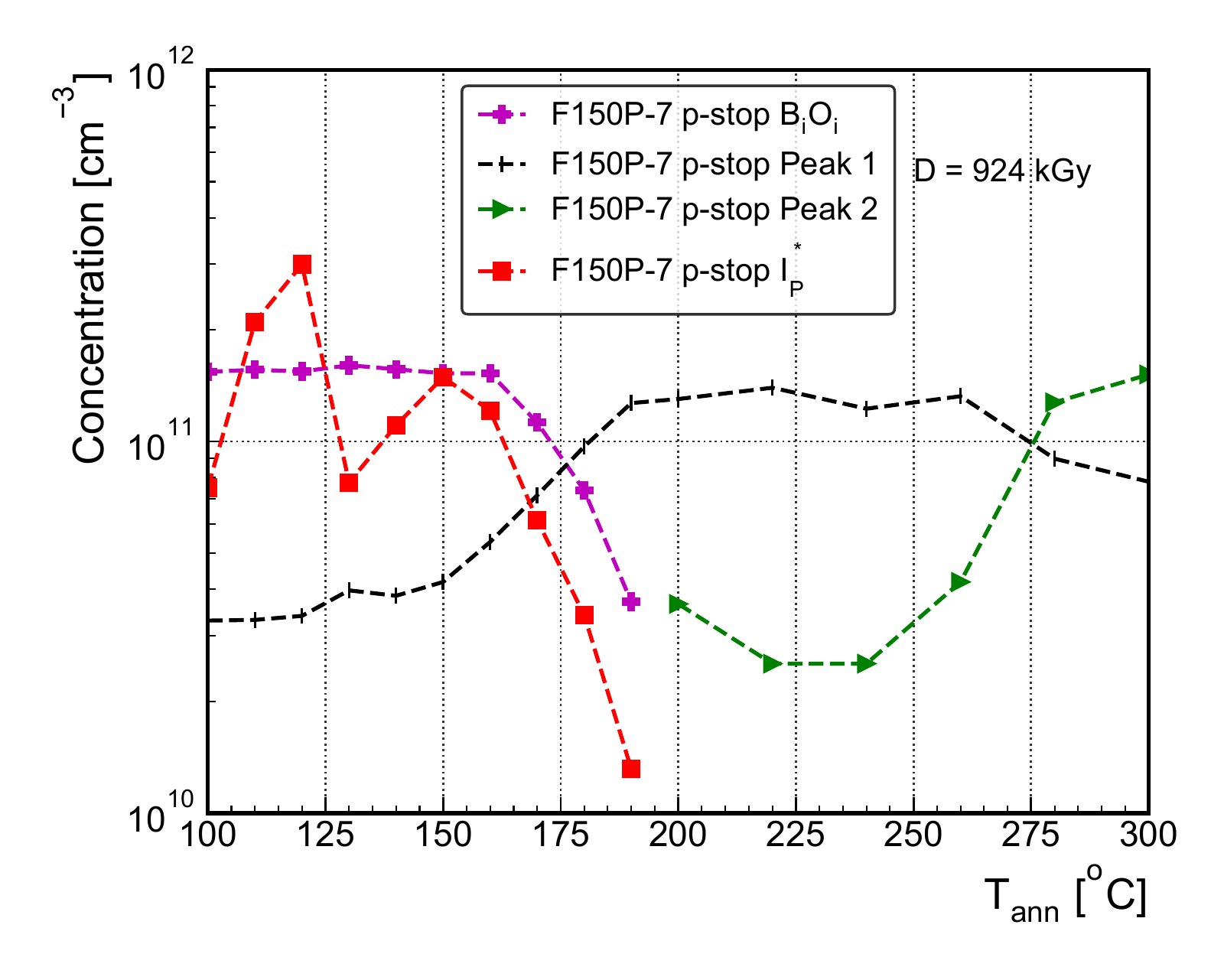}
      }
    \subfloat[\label{fig7d}]{
   \includegraphics[width=0.49\linewidth]{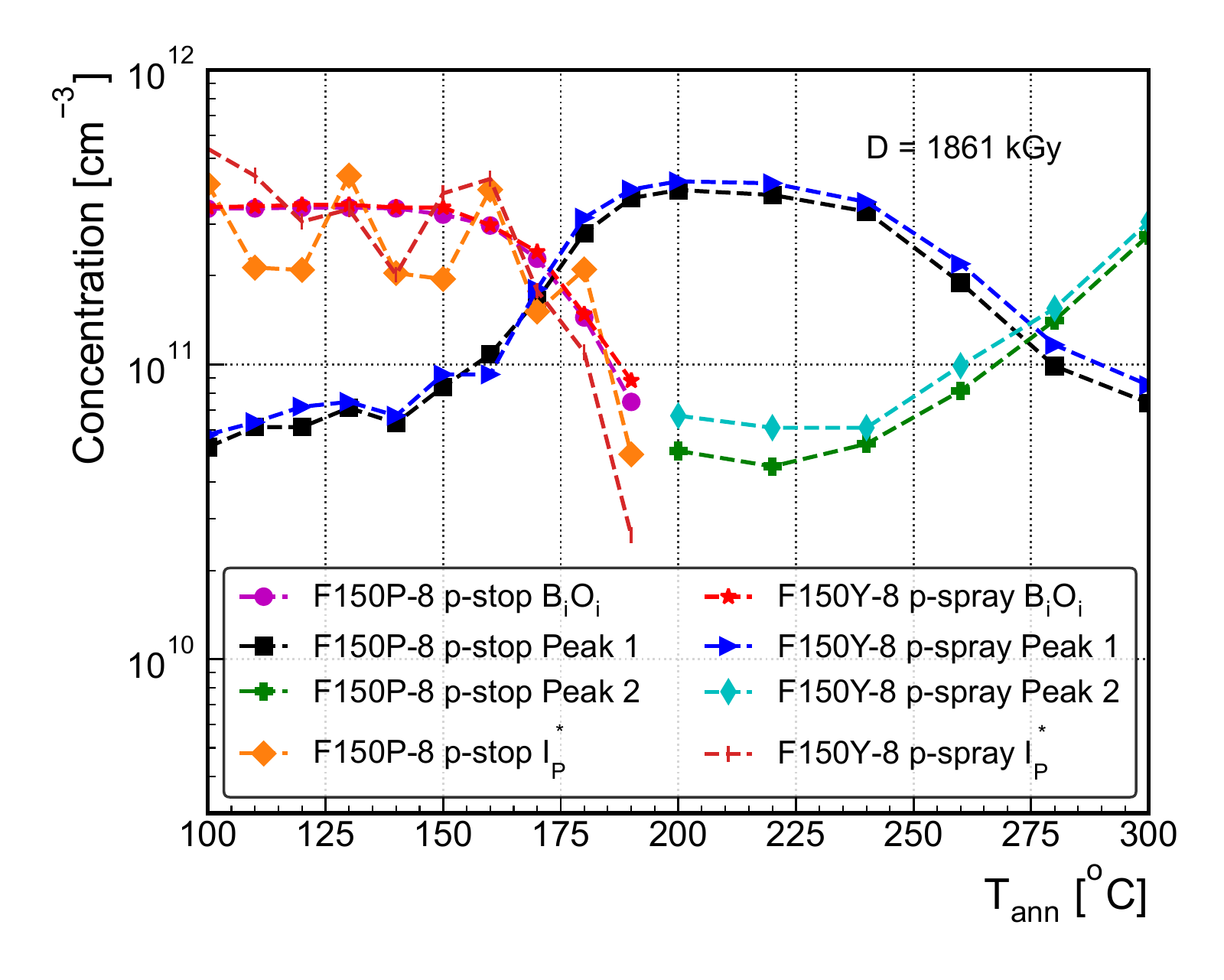}
      }
\caption{(a) TSC spectra for diode F150P-8 after isochronal annealing from \SI{100}{\celsius} to \SI{200}{\celsius}. (b) TSC spectra for diode F150P-8 after isochronal annealing from \SI{200}{\celsius} to \SI{300}{\celsius}. Measurement details for (a) and (b): Trap filling by forward current ($I_{fill}$ = 1mA) injection at $T_{fill}$ = 30~K. The diode was irradiated with $^{60}$Co-$\gamma$ to a dose value of about ${D}$ = $2~\si{MGy}$, and the heating bias voltage was $V_{bias}$ = -300~V. (c) Evolution of the defects (B$_\text{i}$O$_\text{i}$, Peak 1 and Peak 2) concentrations in diode F150P-7 as a function of annealing temperature $T_\text{ann}$. (d) Evolution of the defects (B$_\text{i}$O$_\text{i}$, Peak 1 and Peak 2) concentrations in diodes F150P-8 and F150Y-8 as a function of annealing temperature $T_\text{ann}$. Both (c) and (d) are determined from TSC measurements in the temperature range of 70~K to 110~K.}
\label{fig7}
\end{figure*}

The TSC spectra recorded on the diode F150P-8 after annealing at different temperatures are given in Fig.~\ref{fig7a} and Fig.~\ref{fig7b}. As it can be seen while H40K anneals out at \SI{120}{\celsius}, B$_{\text{i}}$O$_{\text{i}}$ and I$_\text{p}*$ defects starts to anneal at \SI{160}{\celsius} and are disappearing after the treatment at \SI{200}{\celsius}. Worth noting is the significantly lower thermal stability of the I$_\text{p}*$ defect compared with the previously reported one for the I$_\text{p}$ defect in~\cite{b25}. Thus, although both defects show a quadratic dose dependence, there are significant differences between them concerning trapping parameters (see Table~\ref{tab:table2}) and thermal stability, indicating that they are different structural point defects. The dominant C$_{\text{i}}$O$_{\text{i}}$ defect is stable in all the studied temperature ranges.

\par

A further interesting observation is, that the decrease in the concentration of B$_{\text{i}}$O$_{\text{i}}$ and I$_\text{p}^*$ defects are accompanied by the formation of another center ($defect 1$) of which concentration is growing up to about \SI{200}{\celsius} (see Fig.~\ref{fig7a}). After dissociation of B$_{\text{i}}$O$_{\text{i}}$ defect, boron atom return on the substitutional  site (B$_{\text{s}}$) and is recovering its acceptor character.  This process is reflected in an increase of $N_{\text{eff}}$ with twice the amount of the dissociated B$_{\text{i}}$O$_{\text{i}}$ donor. For the $p$-spray diode, the variation in the $N_{\text{eff}}$ given in Fig.~\ref{fig7b} is [$\Delta N_{\text{eff}}$]~=~$\num{7.25e11}$~$\si{cm^{-3}}$ (between \SI{140}{\celsius} and \SI{200}{\celsius}). The total amount of dissociated B$_{\text{i}}$O$_{\text{i}}$ defect for the same sample is [B$_{\text{i}}$O$_{\text{i}}$]~=~$\num{3.5e11}$~$\si{cm^{-3}}$ (see Fig.~\ref{fig7d}), almost a half of [$\Delta N_{\text{eff}}$]. Because the initial Boron doping is mostly restored during the dissociation of B$_{\text{i}}$O$_{\text{i}}$ defect and considering the concentration of the newly formed $defect 1$ (of $\num{4.2e11}$~$\si{cm^{-3}}$), we conclude that the Boron atom cannot be part of $defect 1$.
\par
\begin{figure}[!htb]
\centering
\includegraphics[width=1.0\linewidth]{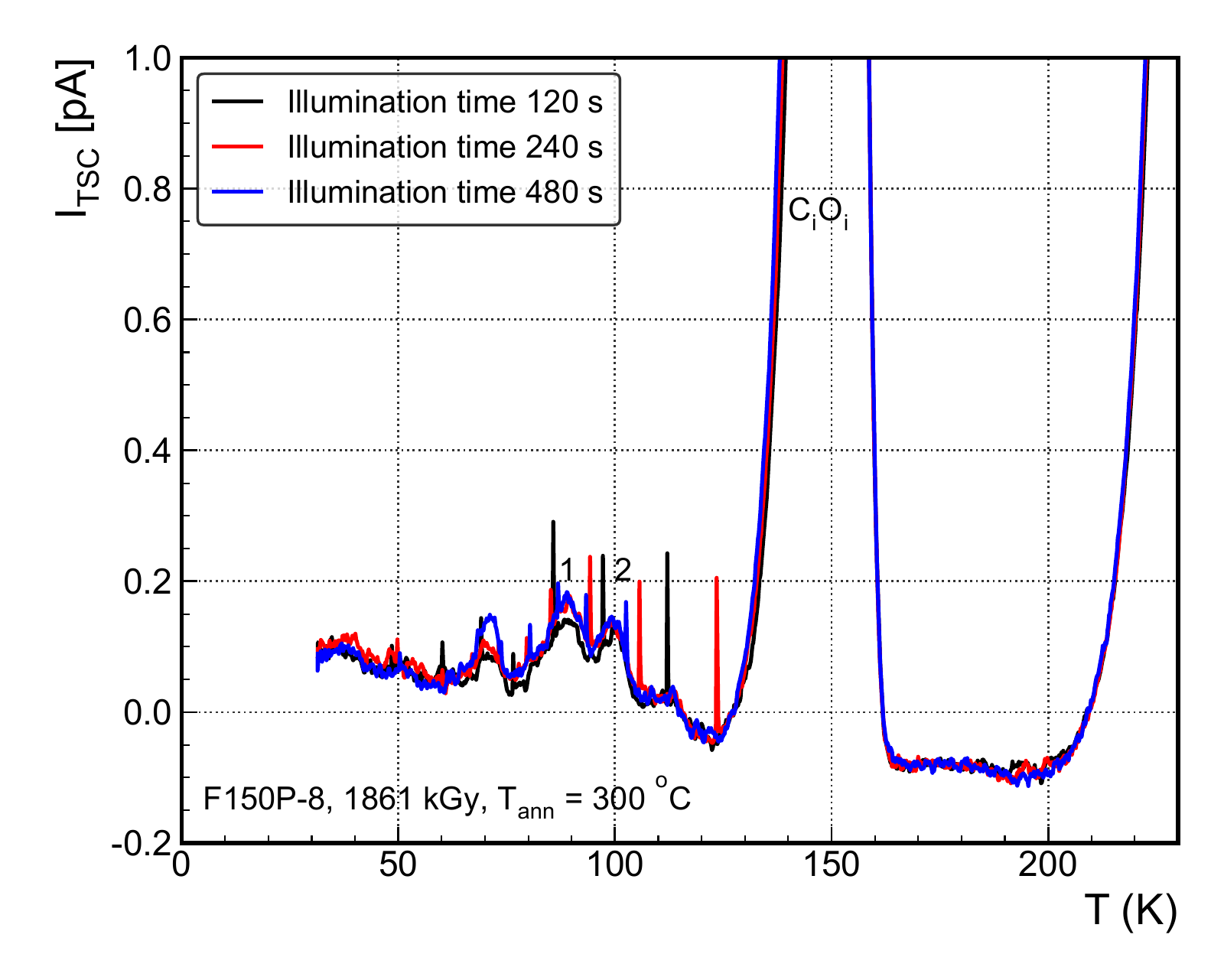}
\caption{Light injection performed on diode F150P-8 after isochronal annealing at $T_\text{ann}$ = 300~$^o$C. Measurement details: Illumination time $t_\text{fill}$ = 120, 240 and 480~s, $V_\text{bias}$ = -200V.}
\label{fig11}
\end{figure}

The $defect 1$ is stable up to about \SI{240}{\celsius} (see Fig.~\ref{fig7b}) when starts to transform in $defect 2$ up to the limit of the used oven (\SI{300}{\celsius}). The variation in defect concentration is given in Fig.~\ref{fig7c} and Fig.~\ref{fig7d}. From TSC experiments with different applied bias voltages, no change in the position of the peak is found for defects 1 and 2. Also, no change in the $N_{\text{eff}}$ values are determined during the annealing transformations occurring between \SI{200}{\celsius} and \SI{300}{\celsius}. This indicates that both defects (1 and 2) are in a neutral charge state at room temperature. In addition, experiments for injecting only holes at $T_\text{fill}$ were performed. For this, a 520~nm wavelength light was used to illuminate the front side (n+) of an over depleted diode at $T_\text{fill}$. Since the absorption length of 520 nm light is roughly \SI{1}{\micro \meter}~\cite{b36, b37}, it can be assumed that only holes are injected into the bulk of reverse biased illuminated diode. The defects 1 and 2 are both detected in this way (see Fig.~\ref{fig11}), evidencing that they are acting as traps for holes.  

\subsection{Surface effect}\label{sec:3.3}

\begin{figure*}[!htp]
 \centering
  \subfloat[\label{fig8a}]{
   \includegraphics[width=0.49\linewidth]{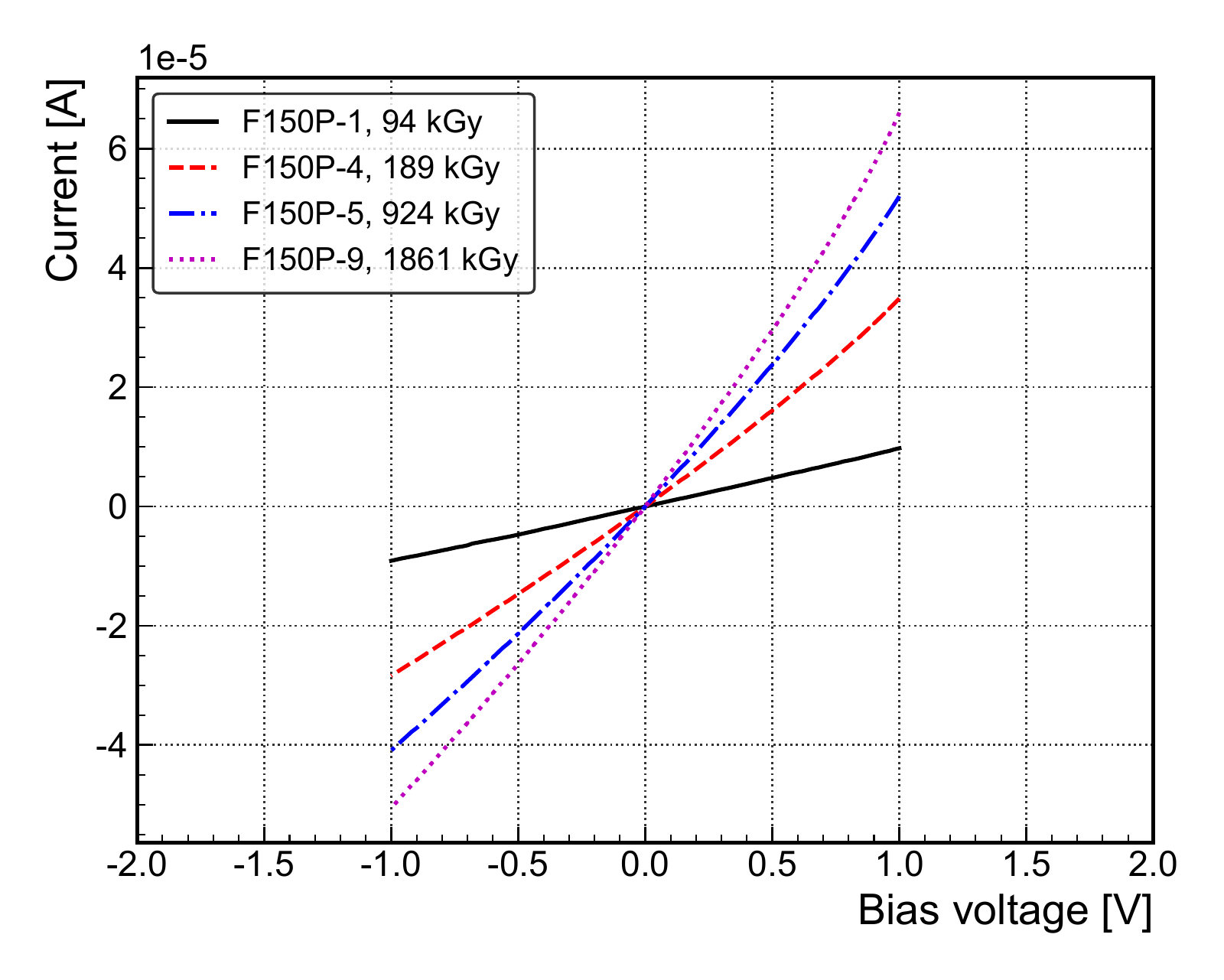}
   }
   \subfloat[\label{fig8b}]{
   \includegraphics[width=0.49\linewidth]{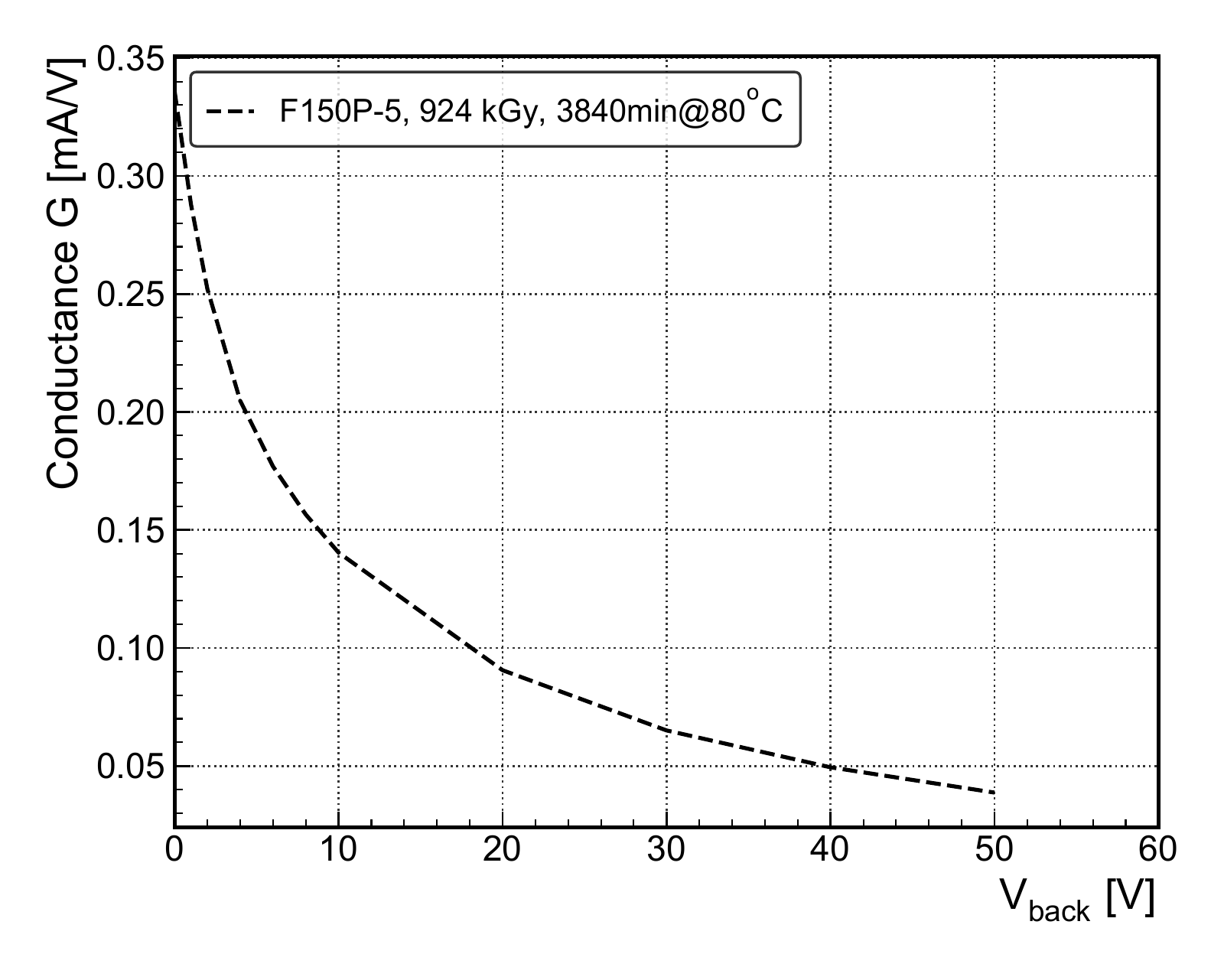}
      }\\
   \subfloat[\label{fig8c}]{
   \includegraphics[width=0.49\linewidth]{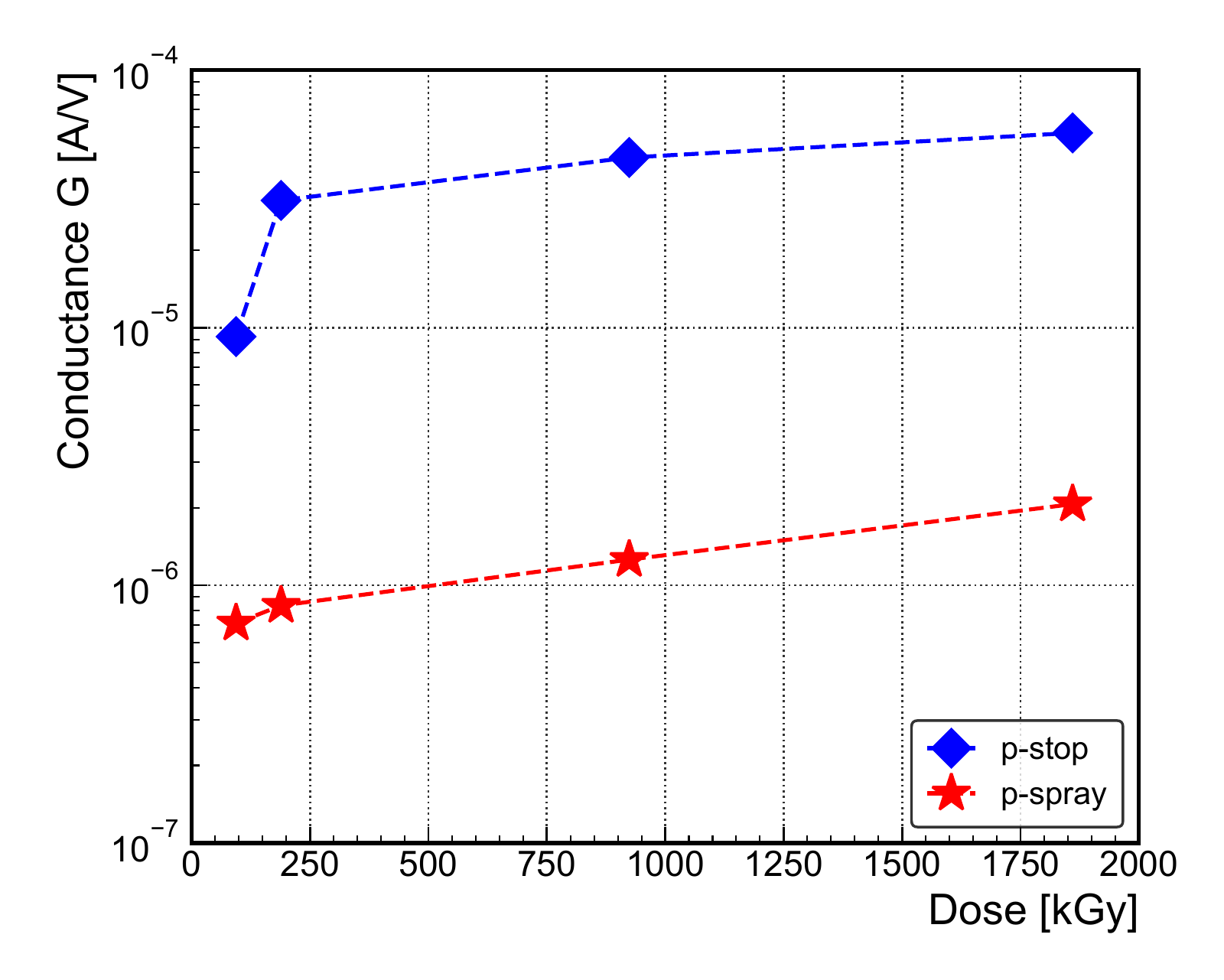}
      }
\caption{(a) Current measured on the pad with varies bias on the guard from -1~V to 1~V for $p$-stop diodes with different irradiation dose values, $V_\text{back}$ = -10~V. (b) Conductance $G$ developed with bias on the backside while grounding the guard ring. (c) The development of $G$ with dose, $V_\text{back}$ = -10~V.}
\label{fig8}
\end{figure*}
\begin{figure*}[!htb]
 \centering
  \subfloat[\label{fig9a}]{
   \includegraphics[width=0.49\linewidth]{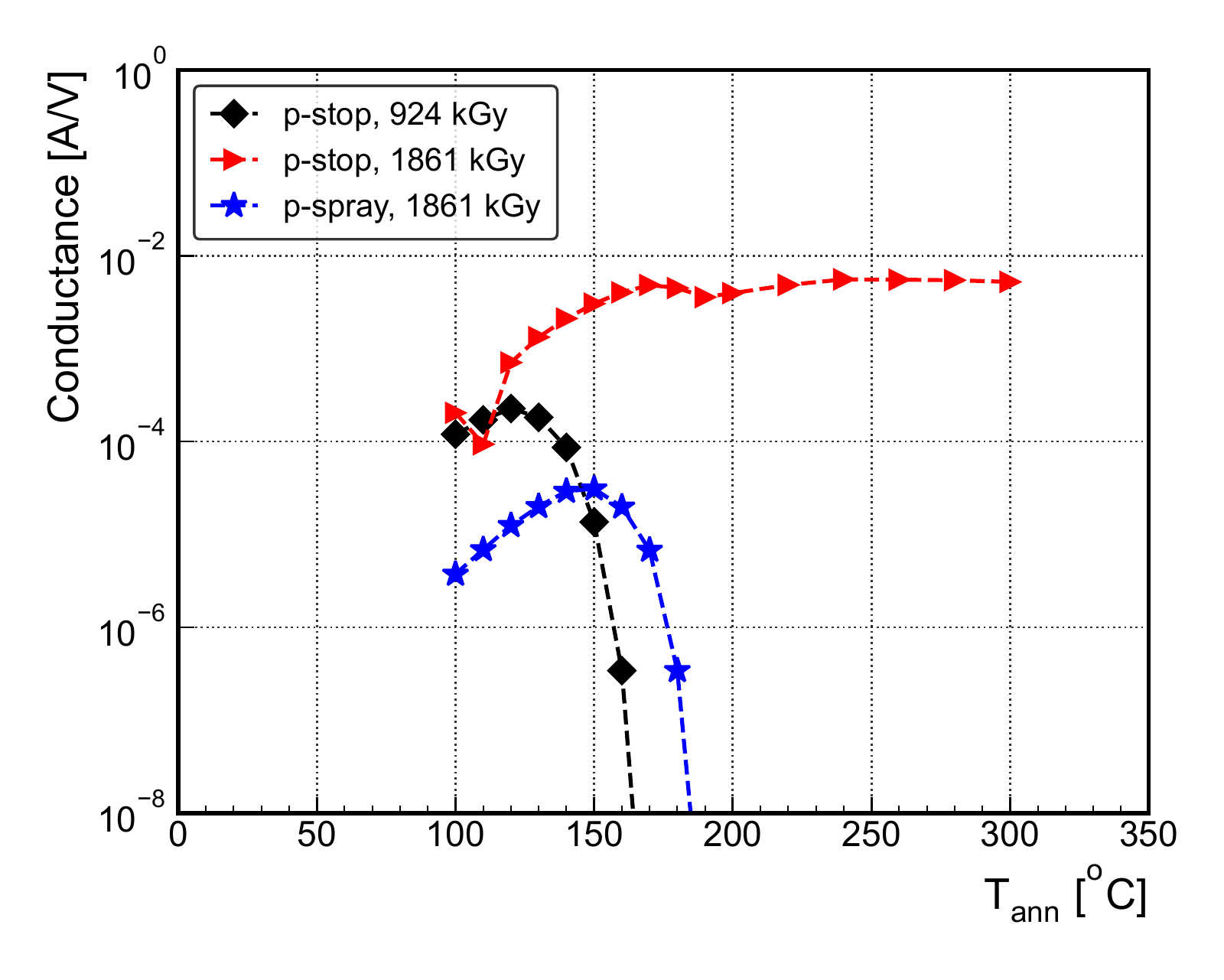}
   }
   \subfloat[\label{fig9b}]{
   \includegraphics[width=0.49\linewidth]{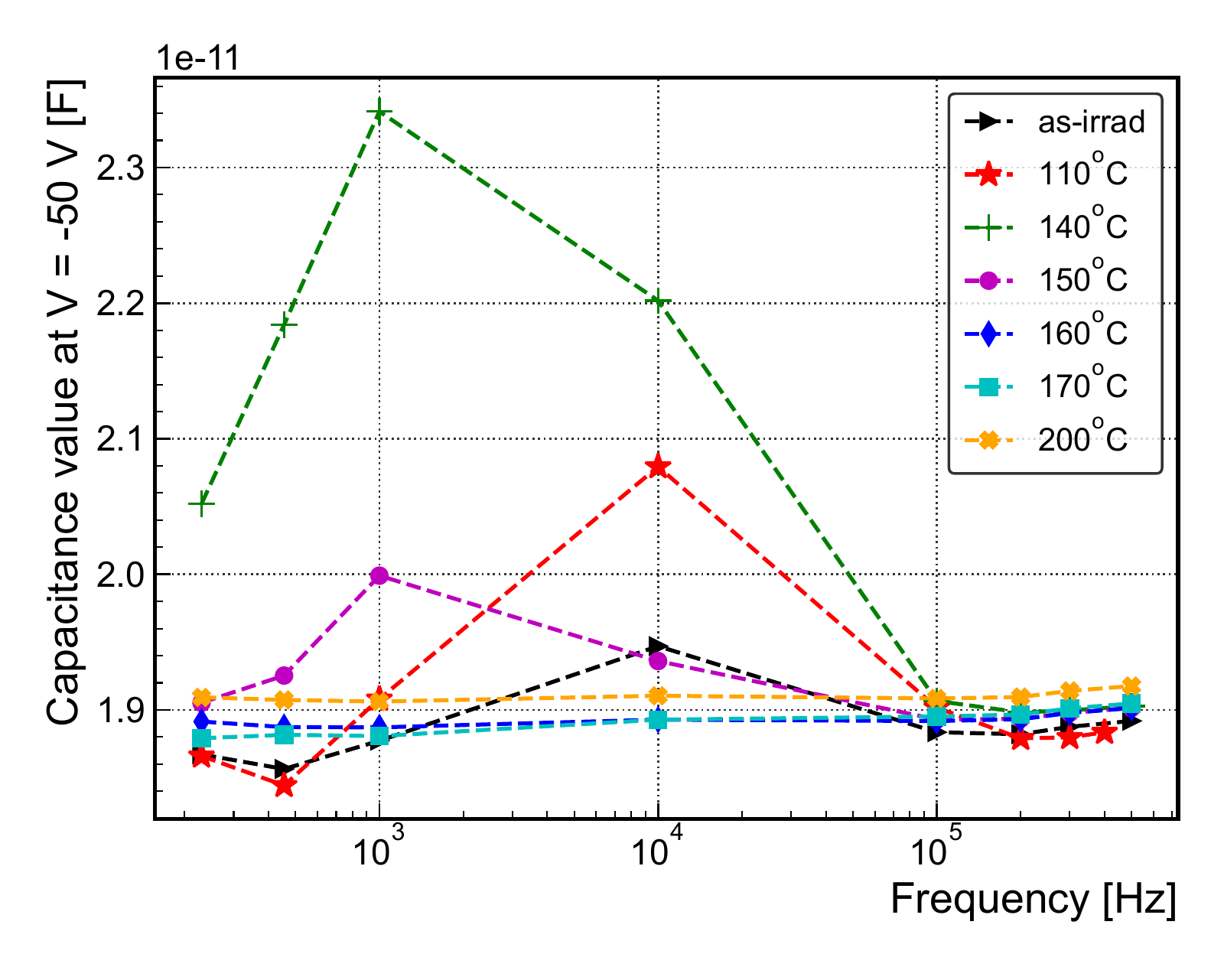}
      } 

\caption{(a) Development of surface conductance with $T_\text{ann}$ at backside applied -10~V. (b) Capacitance value when applied V~=~-50~V vs.\ frequency developed with annealing for $p$-stop after irradiation with $D$ = 924~kGy.}
\label{fig9}
\end{figure*}

Contrary to the investigated $p$-spray diodes, the $p$-stop devices after high dose values (1~MGy and 2~MGy) show a frequency dependence of the $C$--$V$ characteristics, an important source of errors when estimating $N_{\text{eff}}$ depth profiles. This was shown previously for the 230~Hz $\sim$ 10~kHz frequency range~\cite{b35}. As we will show further, the phenomena became more pronounced after high annealing temperatures. Also, the $I$--$V$ characteristics change their behaviour, e.g.\ the pad current at low bias voltages became approximately zero while the current of the guard ring became large at zero voltage and decreases with increasing bias voltage. However, the same behaviours do not appear on the $p$-spray diodes. These findings led to the speculation that surface effects occurring between the guard ring and the pad area might be responsible for the above mentioned observations. In order to study such surface effects, the standard $I$--$V$ setup was slightly modified compared to Fig.~\ref{fig10c} and Fig.~\ref{fig10d}. While the guard ring was connected to the ground via a voltage source, the pad was connected to a current meter. In addition, a second voltage source was connected to the backside of the diode (reverse bias $V_\text{back}$). With this setup, the current flow between the guard ring and the pad could be measured by varying the voltage of the guard ring in the range from -1~V to +1~V. At the same time, a reverse voltage was applied on the backside. As an example, Fig.~\ref{fig8a} presents such surface $I$--$V$ curves at a constant $V_\text{back}$~=~10~V for several $p$-stop diodes irradiated to different dose values.

\par

The slope of the surface current in the linear range ($-$0.5~V to $+$0.5~V) provides the surface conductance $G$, given by:
\begin{eqnarray}
G & = & {{dI} \over {dV}}
\end{eqnarray}
Figure~\ref{fig8b} demonstrates the dependence of the conductance $G$ versus $V_\text{back}$. The decrease of $G$ with increasing V$_\text{back}$ indicates a decrease in the charge carrier concentration near the surface. For $V_\text{back}$ $\geq$ 50~V the diode F150P-5 irradiated to 0.924~MGy becomes fully depleted. The dose dependence of the surface conductance is plotted in Fig.~\ref{fig8c}, for both, $p$-stop and $p$-spray diodes. There is more than one order of magnitude difference between the $G$ values in the two types of diodes. This indicates that a much larger interface trap concentration exists in the $p$-stop diodes compared to the $p$-spray ones. 
\par
Isochronal annealing studies were performed for two $p$-stop diodes (F150P-7 (0.924~MGy), F150P-8 (1.861~MGy)) and one $p$-spray diode (F150Y-8 (1.861~MGy)). 

\par
The dependence of $G$ on the annealing temperature $T_\text{ann}$ is shown in Fig.~\ref{fig9a}. For the 924~kGy irradiated diodes, the changes in $G$ are observed first as an increase for annealing up to \SI{120}{\celsius} and \SI{150}{\celsius}, respectively, followed by a rapid decrease with increasing $T_\text{ann}$. At the same time, the conductance of the $p$-stop sample is much larger than that of the $p$-spray one. On the other hand, in the highly irradiated $p$-stop device, $G$ increases up to about \SI{120}{\celsius} and stays thereafter nearly constant up to the end of the annealing experimental, at the temperature of \SI{300}{\celsius}. A frequency dependence in $C$--$V$ measurements is only observed if $G$ is larger than $\num{3e-6}$ or $\num{2e-6}$~A/V at $V_{back}$ = 10 or 50~V, respectively. The frequency dependencies of the capacitance measured after annealing at different temperatures on F150P-7 $p$-stop diode are shown in Fig.~\ref{fig9b} in the 230~Hz to 500~kHz frequency range. The capacitance values are taken for a reverse bias of 50~V, which is slightly below the full depletion voltage. As can be seen, the maximal capacitance value is observed for $T_\text{ann}$ = \SI{140}{\celsius} and a frequency of 1 kHz. From \SI{160}{\celsius} on, the frequency dependence vanishes, in accordance with the corresponding very low conductance shown in Fig.~\ref{fig9a}. It is worth noting that, above 100~kHz, the capacitance is constant for all the annealing temperatures. Measuring the capacitance in such a frequency-independent range is considered to be a safe procedure for performing the $C$--$V$ measurements and extracting further the $N_\text{eff}$ profiles. Under this assumption, the errors for all capacitance values at 500~kHz are below 4\% for $p$-stop diodes with dose values below 1~MGy. Due to the conductance after 1.861~MGy doses, the capacitance of $p$-stop is not reliable anymore after annealing.

\par
The errors in the determined $N_{\text{eff}}$ in Fig.~\ref{fig6b} caused by averaging range are around 1\% (varying average ranges in between \SI{60}{\micro \meter} and \SI{120}{\micro \meter}). The main errors of $N_{\text{eff}}$ are from the frequency-dependent $C$--$V$ measurements, which will be discussed in detail in subsection~\ref{sec:3.3}. 

\section{Conclusion} \label{sec:4}
The radiation damage of two types of silicon diodes ($p$-spray and $p$-stop), manufactured on $p$-type Fz-material, with a resistivity of about 4~k$\Omega$cm and exposed to $^{60}$Co $\gamma$-ray at different dose values (0.1~MGy, 0.2~MGy, 1~MGy, and 2~MGy) was investigated in this study. $I$--$V$ and $C$--$V$ measurements were employed for studying the radiation-induced changes in the densities of leakage current ($J_d$) and effective space charge ($N_\text{eff}$). The results showed that with increasing the dose, $N_\text{eff}$ decreases while $J_d$ linearly increases, the latter exhibiting a similar trend to that of oxygen enriched $n$-type silicon irradiated with $^{60}$Co $\gamma$-ray. In order to identify the $\gamma$-induced traps in the bulk of the samples, TSC measurements were performed on irradiated $p$-type diodes. The TSC spectra revealed the presence of defects, including H40K, VO, B$_\text{i}$O$_\text{i}$, C$_\text{i}$O$_\text{i}$, $V_2$, and the newly detected I$_\text{P}^*$. The results showed that all the concentrations of defects increased linearly with dose except the I$_\text{P}^*$ that shows a quadratic dose dependence. The process of multi-phonon capture of holes by the C$_\text{i}$O$_\text{i}$ defect was investigated by analyzing the corresponding TSC peak obtained after the filling of the defect was performed at different temperatures. The obtained results differ from those determined after irradiation with 1~MeV neutrons of $n$-type silicon diode, the capturing process on the C$_\text{i}$O$_\text{i}$ defect being faster after irradiation with $\gamma$-rays. This is explained considering that there are additional potential barriers surrounding the disordered (clustered) regions produced by hadron which slow down the charge transport in the diodes.
\par
To gain a better understanding of the defects' thermal stability, kinetics and their impact on the device properties, isochronal annealing experiments were performed in the \SI{100}{\celsius}-\SI{300}{\celsius} temperature range. The TSC experiments revealed that C$_\text{i}$O$_\text{i}$ complex is thermally stable up to the end of the isochronal annealing, performed up to \SI{300}{\celsius} in our study. In contrast, the other radiation-induced defects change their concentrations during the thermal treatment. Thus, H40K anneals out at \SI{120}{\celsius} while B$_\text{i}$O$_\text{i}$ and I$_\text{P}^*$ start to anneal out after annealing at \SI{140}{\celsius} and the devices are recovering most of their initial doping. Two unidentified traps are detected to form during the annealing, labelled as $defect 1$ and $defect 2$. $defect 1$ is forming during the treatment in the \SI{140}{\celsius}-\SI{200}{\celsius} temperature range and transforms into $defect 2$ after annealing at higher temperatures. Both defects were found to be hole traps and neutral charged at room temperature.
\par
The results of the macroscopic measurements showed that the density of leakage current ($J_d$) measured in full depletion conditions (at $V$~=~-300~V) remains stable for annealing temperatures up to \SI{200}{\celsius}, indicating that none of the annealed out defects has an influence on the measured current. In the \SI{200}{\celsius}-\SI{300}{\celsius} temperature range of annealing, $J_d$ increases up to \SI{260}{\celsius} and then decreases. No evidence for bulk defects responsible for such behaviour was detected in TSC experiments. Variations of $N_\text{eff}$ were observed only in the \SI{150}{\celsius}-\SI{200}{\celsius} temperature annealing range, where the dissociation of the B$_\text{i}$O$_\text{i}$ defect occurs and $N_\text{eff}$ increases because the initial removed substitutional Boron is recovered. Depending on the isolation technique used for the fabrication of the two types of investigated diodes, different frequency dependences were evidenced, according not only to the irradiation dose but also to the isolation technique used for separating the pad from the guard ring. To further investigate this phenomenon, surface conductance ($G$) measurements were performed on different samples. The frequency dependence in $C$--$V$ measurements is only observed if $G$ was large than $\num{3e-6}$ or $\num{2e-6}$~A/V at $V_\text{back}$ = 10 or 50~V, respectively. This is not the case for most of the irradiated $p$-spray diodes, apart from this highest dose of 1861 kGy after annealing at high temperature. For irradiated $p$-stop diodes, the value of $G$ is consistently above this limit, but it can be decreased by annealing for $D$ value below 1 MGy (F150P-7).

\section*{Acknowledgment}
This work has been carried out in the framework of the RD50 Collaboration. The project has received funding from the European Unions Horizon 2020 Research and Innovation program under Grant Agreement no.\ 654168. C.~Liao would like to thank the Deutsche Forschungsgemeinschaft (DFG, German Research Foundation) under Germany's Excellence Strategy -- EXC2121 "Quantum Universe" -- 390833306 and Professor Z.~Li for supporting his stay at the University of Hamburg. I.~Pintilie acknowledges the funding received through IFA-CERN-RO 8/2022 project. Z.~Li acknowledges the funding received through the Key Scientific and Technological Innovation Project of Shandong Province under Grant No. 2019 TSLH 0316, and the project of Yantai Institute for the exchange of Driving Forces under Grants No. 2019XJDN002.

\bibliographystyle{elsarticle-num-names}

\begin{thebibliography}{00}
\bibitem{b1} G. Kramberger et al., ``Radiation effects in Low Gain Avalanche
Detectors after hadron irradiations,'' \textit{J. Instrum.}, vol. 10, no. 7, p.
P07006, 2015.
DOI:
\href{https://iopscience.iop.org/article/10.1088/1748-0221/10/07/P07006}{10
.1088/1748-0221/10/07/p07006}

\bibitem{b2} M. Ferrero et al., ``Radiation resistant LGAD design,''
\textit{Nucl. Instrum. Methods Phys. Res. A}, vol. 919, pp. 16--26, 2019.
DOI:
\href{https://doi.org/10.1016/j.nima.2018.11.121}{https://doi.org/10.1016/j
.nima.2018.11.121}

\bibitem{b3} P. M. Mooney, L. J. Cheng, M. S\"uli, J. D. Gerson, and J. W.
Corbett, ``Defect energy levels in boron-doped silicon irradiated with 1-MeV
electrons,'' \textit{Phys. Rev. B}, vol. 15, no. 8, pp. 3836--3843, 1977.
DOI:
\href{https://journals.aps.org/prb/abstract/10.1103/PhysRevB.15.3836}{10.1103
/PhysRevB.15.3836}

\bibitem{b4} I. Pintilie, E. Fretwurst and G. Lindstr\"om, ``Cluster related
hole trap with enhanced-field-emission the source for long term annealing in
hadron irradiated Si diodes,'' \textit {Appl. Phys. Lett.}, vol. 92, p. 024101,
2008.
DOI:
\href{https://aip.scitation.org/doi/full/10.1063/1.2832646}{10.1063/1.2832646}

\bibitem{b5} Kh. A. Abdullin, B. N. Mukashev and  Yu. V. Gorelkinskii, ``Metastable oxygen - silicon interstitial complex in crystalline silicon,'' \textit {Semicond. Sci. Technol.}, vol. 11, no. 11, 1996.
DOI:
\href{https://dx.doi.org/10.1088/0268-1242/11/11/010}{10.1088/0268-1242/11/11/010}

\bibitem{b22} I. Pintilie, G. Lindstr\"om, A. Junkes, and E. Fretwurst, "Radiation-induced point- and cluster-related defects with strong impact on damage properties of silicon detectors," \textit{Nucl. Instrum. Methods Phys. Res. A}, vol. 611, no.1, pp. 52-68, 2009.
DOI: \href{https://www.sciencedirect.com/science/article/pii/S0168900209018129}{https://doi.org/10.1016/j.nima.2009.09.065}

\bibitem{b6} J. Coutinho et al., ``Electronic and dynamical properties of the silicon trivacancy,'' \textit{Phys. Rev. B}, vol. 86, p. 174101, Nov. 2012. 
DOI: \href{https://link.aps.org/doi/10.1103/PhysRevB.86.174101}{10.1103/PhysRevB.86.174101}

\bibitem{b33} L. C. Kimerling, M. T. Asom,  J. L. Benton, P. J. Drevinsky, and
C. E. Caefer, ``Interstitial defect reactions in silicon,'' In \textit{Defects
in Semiconductors 15}, vol. 38 in \textit{Materials Science Forum}, pp.
141--150, 1989.
DOI:
\href{https://doi.org/10.4028/www.scientific.net/MSF.38-41.141}{https://doi.org
/10.4028/www.scientific.net/MSF.38-41.141}


\bibitem{b38} C. M\"oller, and K. Lauer, ``Light-induced degradation in indium-doped silicon,'' \textit{Phys. Status Solidi RRL}, vol. 7, no. 7, pp. 461--464, 2013. 
DOI:
\href{https://onlinelibrary.wiley.com/doi/abs/10.1002/pssr.201307165}{https://doi.org/10.1002/pssr.201307165}

\bibitem{b39} K. Lauer et al., ``Activation energies of the In$_\text{Si}$-Si$_\text{i}$ defect transitions obtained by carrier lifetime measurements,'' \textit{Phys. Status Solidi C}, vol. 14, no. 5, pp. 1600033, 2017. 
DOI:
\href{https://onlinelibrary.wiley.com/doi/abs/10.1002/pssc.201600033}{https://doi.org/10.1002/pssc.201600033}


\bibitem{b7} Y. Gurimskaya et al., ``Radiation damage in p-type EPI silicon pad diodes irradiated with protons and neutrons,'' \textit{Nucl. Instrum. Methods Phys. Res. A.}, vol. 958, p. 162221, 2020.
DOI:
\href{https://www.sciencedirect.com/science/article/pii/S0168900219307181}{https://doi.org/10.1016/j.nima.2019.05.062}

\bibitem{b8} C. Liao et al., ``The Boron Oxygen ($\si{B_i O_i}$) Defect
Complex Induced by Irradiation with 23 GeV Protons in p-Type Epitaxial Silicon
Diodes,'' \textit{IEEE Trans. Nucl. Sci.}, vol. 69, no. 3, pp. 576--586, Mar
2022. 
DOI:
\href{https://ieeexplore.ieee.org/document/9698047}{10.1109/TNS.2022.3148030}

\bibitem{b9} A. Himmerlich et al., "Defect characterization studies on neutron irradiated boron-doped silicon pad diodes and Low Gain Avalanche Detectors," \textit{Nucl. Instrum. Methods Phys. Res. A.}, vol. 1048, p. 167977, 2023.
DOI:
\href{https://www.sciencedirect.com/science/article/pii/S0168900222012694}{https://doi.org/10.1016/j.nima.2022.167977}

\bibitem{b10} C. Liao et al., ``Investigation of the Boron removal effect induced by 5.5 MeV electrons on highly doped EPI- and Cz-silicon,'' \textit{arXiv:2306.14736}, physics. app-ph, 2023.
DOI:
\href{https://arxiv.org/abs/2306.14736}{https://doi.org/10.48550/arXiv.2306.14736}

\bibitem{b11} C. Besleaga, A. Kuncser, A. Nitescu, G. Kramberger, M. Moll and I. Pintilie, ``Bistability of the BiOi complex and its implications on evaluating the "acceptor removal" process in p-type silicon,'' \textit{Nucl. Instrum. Methods Phys. Res. A}, vol. 1017, p. 165809, 2021.
DOI: \href{https://www.sciencedirect.com/science/article/pii/S0168900221007944}{https://doi.org/10.1016/j.nima.2021.165809}

\bibitem{b12} J. H. Cahn, ``Irradiation Damage in Germanium and Silicon due to Electrons and Gamma Rays,'' \textit {J. Appl. Phys.}, vol. 30, p. 1310-1316,
1956.
DOI:
\href{https://doi.org/10.1063/1.1735310}{10.1063/1.1735310}

\bibitem{b13} M. Moll, ``Radiation damage in silicon particle detectors: Microscopic defects and macroscopic properties,'' PhD dissertation, Dept. Phys., Univ. Hamburg, Hamburg, Germany, 1999. \href{https://bib-pubdb1.desy.de/record/300958}{DESY-THESIS-1999-040}

\bibitem{b40} \href{https://www.hamamatsu.com/eu/en.html}{https://www.hamamatsu.com/eu/en.html}

\bibitem{b14} The Tracker Group of the CMS Collaboration, "Evaluation of HPK $n^+$-$p$ planar pixel sensors for the CMS Phase-2 upgrade," \textit {Nucl. Instrum. Methods Phys. Res. A.}, vol. 1053, p. 168326, 2023. 
DOI:\href{https://www.sciencedirect.com/science/article/pii/S0168900223003169}{https://doi.org/10.1016/j.nima.2023.168326}

\bibitem{b15} J. Schwandt, ``CMS Pixel detector development for the HL-LHC,'' \textit {Nucl. Instrum. Methods Phys. Res. A.}, vol. 924, p. 59-63, 2019.
DOI:\href{https://www.sciencedirect.com/science/article/pii/S0168900218310805}{https://doi.org/10.1016/j.nima.2018.08.121}

\bibitem{b16} M. Roguljic, A. Starodumov, A. Karadzhinova-Ferrer, D. Ferencek, A. A. Ahmed and L.M. Jara-Casas, ``Low dose rate $^{60}$Co facility in Zagreb,'' \textit {PoS.}, vol. Vertex2019, p. 066, 2020.
DOI:\href{https://pos.sissa.it/373/066}{10.22323/1.373.0066}


\bibitem{b17} I. Pintilie, L. Pintilie, M. Moll, E. Fretwurst, and G. Lindstr\"om, ``Thermally stimulated current method applied on diodes with high concentration of deep trapping levels,'' \textit {Appl. Phys. Lett.}, vol. 78, p. 550, 2001.
DOI: \href{https://doi.org/10.1063/1.1335852}{10.1063/1.1335852}

\bibitem{b18} I. Pintilie, M. Buda, E. Fretwurst, G. Lindstr\"om, and J. Stahl, ``Stable radiation-induced donor generation and its influence on the radiation tolerance of silicon diodes,'' \textit{Nucl. Instrum. Methods Phys. Res. A}, vol. 556, no.1, pp. 197--208, 2006.
DOI: \href{https://doi.org/10.1016/j.nima.2005.10.013}{https://doi.org/10.1016/j.nima.2005.10.013}

\bibitem{b19} \href{https://www.thermofisher.com/de/de/home/life-science/lab-equipment/lab-ovens-furnaces/lab-heating-drying-ovens.html}{https://www.thermofisher.com/de/de/home/life-science/lab-equipment/lab-ovens-furnaces/lab-heating-drying-ovens.html}

\bibitem{b20} E. Fretwurst et al., ``Bulk damage effects in standard and oxygen-enriched silicon detectors induced by $^{60}$Co-gamma radiation,'' \textit{Nucl. Instrum. Methods Phys. Res. A}, vol. 514, no.1, pp. 1-8, 2003.
DOI: \href{https://www.sciencedirect.com/science/article/pii/S0168900203023660}{https://doi.org/10.1016/j.nima.2003.08.077}

\bibitem{b21} A. Hall\'en, N. Keskitalo, F. Masszi, and V. N\'agl, ``Lifetime in proton irradiated silicon,'' \textit{J. Appl. Phys.}, vol. 79, no.8, pp. 3906-3914, 1996.
DOI: \href{https://aip.scitation.org/doi/abs/10.1063/1.361816}{10.1063/1.361816}


\bibitem{b23} R. Radu, I. Pintilie, L. C. Nistor, E. Fretwurst, G. Lindstr\"om,
and L. F. Makarenko, ``Investigation of point and extended defects in electron
irradiated silicon Dependence on the particle energy,'' \textit{J. Appl.
Phys.}, vol. 117, p. 164503, 2015. 
DOI: \href{https://doi.org/10.1063/1.4918924}{10.1063/1.4918924}

\bibitem{b24}  I. Pintilie, E. Fretwurst, G. Lindstr\"om, and J. Stahl, "Second-order generation of point defects in gamma-irradiated float-zone silicon, an explanation for "type inversion"," \textit{Appl. Phys. Lett.}, vol. 82, no. 13, pp. 2169--2171, 2003. 
DOI: \href{https://doi.org/10.1063/1.1564869}{10.1063/1.1564869}

\bibitem{b25} I. Pintilie, E. Fretwurst, G. Kramberger, G. Lindstr\"om, Z. Li, and J. Stahl, "Second-order generation of point defects in highly irradiated float zone silicon-annealing studies," \textit{Physica B: Condensed Matter.}, vol. 340-342, pp. 578--582, 2003.
DOI: \href{https://www.sciencedirect.com/science/article/pii/S0921452603007804}{https://doi.org/10.1016/j.physb.2003.09.131}

\bibitem{b26} H. Feick, ``Radiation Tolerance of Silicon Particle Detectors for High-Energy Physics Experiments,'' PhD dissertation, Dept. Phys., Univ. Hamburg, Hamburg, Germany, 1997. \href{http://cds.cern.ch/record/335685}{DESY-F-35-D-97-8}

\bibitem{b27} C. T. Sah, L. Forbes, L. L. Rosier, and A. F. Tasch, ``Thermal and optical emission and capture rates and cross sections of electrons and holes at imperfection centers in semiconductors from photo and dark junction current and capacitance experiments,'' \textit{Solid State Electron}, vol. 13, no. 6, pp. 759--788, 1970.
DOI: \href{https://doi.org/10.1016/0038-1101(70)90064-X}{https://doi.org/10.1016/0038-1101(70)90064-X}

\bibitem{b28} M. G. Buehler, ``Impurity centers in pn junctions determined from shifts in the thermally stimulated current and capacitance response with heating rate,'' \textit{Solid State Electron}, vol. 15, no. 1, pp.69--79,  1972.
DOI: \href{https://doi.org/10.1016/0038-1101(72)90068-8}{https://doi.org/10.1016/0038-1101(72)90068-8}

\bibitem{b29}  M. A. Green, ``Intrinsic concentration, effective densities of states, and effective mass in silicon,'' \textit{J. Appl. Phys.}, vol. 67, no. 6, pp. 2944--2954, 1990. 
DOI: \href{https://aip.scitation.org/doi/10.1063/1.345414}{10.1063/1.345414}

\bibitem{b30}  A. Chilingarov, "Temperature dependence of the current generated in Si bulk," \textit{J. Instrum}, vol. 8, no. 10, pp. P10003, 2013.
\par
DOI: \href{https://dx.doi.org/10.1088/1748-0221/8/10/P10003}{10.1088/1748-0221/8/10/P10003}

\bibitem{b31}  P. Br\"aunlich, "Thermally Stimulated Relaxation in Silicon," \textit{Clarendon Press.}, section 2.5, 1979.
\par
Link: \href{https://catalogue.nla.gov.au/Record/2253954}{https://catalogue.nla.gov.au/Record/2253954}

\bibitem{b41} B. G. Svensson, B. Mohadjeri, A. Hall\'en, J. H. Svensson and J. W. Corbett, "Divacancy acceptor levels in ion-irradiated silicon," \textit{Phys. Rev. B}, vol. 43, no. 0, pp. 2292--2298, 1991.
\par
DOI: \href{https://link.aps.org/doi/10.1103/PhysRevB.43.2292}{10.1103/PhysRevB.43.2292}



\bibitem{b42} E. V. Monakhov, J. Wong-Leung, A. Yu. Kuznetsov, C. Jagadish and B. G. Svensson, "Ion mass effect on vacancy-related deep levels in Si induced by ion implantation," \textit{Phys. Rev. B}, vol. 65, no. 9, pp. 245201, 2002.
\par
DOI: \href{https://link.aps.org/doi/10.1103/PhysRevB.65.245201}{10.1103/PhysRevB.65.245201}


\bibitem{b43} P. J. Drevinsky, C. E. Caefer, S. P. Tobin, J.C. Mikkelsen and L. C. Kimerling, ``Influence of oxygen and boron on defect production in irradiated silicon,'' \textit{Mater. Res. Soc. Symp. Proc}, vol. 104, pp. 167--172, 1988.
DOI: \href{https://www.cambridge.org/core/journals/mrs-online-proceedings-library-archive/article/abs/influence-of-oxygen-and-boron-on-defect-production-in-irradiated-silicon/14A697AC047B6C2B74EDD2B6708E7D0B}{10.1557/PROC-104-167}


\bibitem{b44} O. V. Feklisova, N. A. Yarykin, and J. Weber,  ``Annealing kinetics of boron-containing centers in electron-irradiated silicon,'' \textit{Semiconductors}, vol. 47, pp. 228--231, 2013.
DOI: \href{https://doi.org/10.1134/S1063782613020085}{https://doi.org/10.1134/S1063782613020085}


\bibitem{b32} A. Himmerlich et al., "Defect characterization studies and modelling of defect spectra for $^{60}$Co gamma-irradiated epitaxial p-type Si diodes," presented at the
41th RD50 Workshop, Sevilla, Spain, 2022.  
[Online]. Available: https://indico.cern.ch/event/1132520/contributions/5141025/


\bibitem{b36}  W. C. Dash and R. Newman, "Intrinsic Optical Absorption in Single-Crystal Germanium and Silicon at 77~$^\text{o}$K and 300~$^\text{o}$K," \textit{Phys. Rev.}, vol. 99, pp. 1151--1155, 1955. 
DOI: \href{https://journals.aps.org/pr/abstract/10.1103/PhysRev.99.1151}{https://doi.org/10.1103/PhysRev.99.1151}

\bibitem{b37}  R. Braunstein, A. R. Moore, and F. Herman, "Intrinsic Optical Absorption in Germanium-Silicon Alloys," \textit{Phys. Rev.}, vol. 109, pp. 695--710, 1958. 
DOI: \href{https://journals.aps.org/pr/abstract/10.1103/PhysRev.109.695}{https://doi.org/10.1103/PhysRev.109.695}





\bibitem{b34}  K. Lauer, K. Peh, S. Krischok, and S. Rei\ss, "Development of Low-Gain Avalanche Detectors in the Frame of the Acceptor Removal Phenomenon," \textit{Phys. Status Solidi (a).}, vol. 219, no. 17, pp. 2200177, 2022. 
DOI: \href{https://onlinelibrary.wiley.com/doi/abs/10.1002/pssa.202200177}{https://doi.org/10.1002/pssa.202200177}

\bibitem{b35} C. Liao et al., "Investigation of high resistivity p-type FZ silicon diodes after $^{60}$Co - gamma irradiation," presented at the
40th RD50 Workshop, CERN, 2022.  
[Online]. Available: https://indico.cern.ch/event/1157463/



\end{thebibliography}

\end{document}